\documentclass[pdflatex,sn-nature]{sn-jnl}

\usepackage{graphicx}%
\usepackage{multirow}%
\usepackage{amsmath,amssymb,amsfonts}%
\usepackage{amsthm}%
\usepackage{mathrsfs}%
\usepackage[title]{appendix}%
\usepackage{xcolor}%
\usepackage{textcomp}%
\usepackage{manyfoot}%
\usepackage{booktabs}%
\usepackage{algorithm}%
\usepackage{algorithmicx}%
\usepackage{algpseudocode}%
\usepackage{listings}%
\usepackage{subcaption} %
\usepackage{gensymb}
\usepackage{threeparttable}
\usepackage{caption} 
\usepackage{makecell}
\usepackage[style=nature]{biblatex}
\theoremstyle{thmstyleone}%
\theoremstyle{thmstyletwo}%

\theoremstyle{thmstylethree}%

\begin{document}

\title[Article Title]{AI worsens climate change, integrated assessment shows}


\author[1,2]{\fnm{Huiying} \sur{Ye}}\email{yehuiying@hebut.edu.cn}
\equalcont{These authors contributed equally to this work.}

\author[3,4,5,6,7]{\fnm{Richard S.J.} \sur{Tol}}\email{R.Tol@sussex.ac.uk}
\equalcont{These authors contributed equally to this work.}

\author*[8]{\fnm{Fangzhi} \sur{Wang}}\email{fzwang@cityu.edu.mo}
\equalcont{These authors contributed equally to this work.}

\affil[1]{\orgdiv{School of Economics and Management}, \orgname{Hebei University of Technology}, \orgaddress{ \city{Tianjin}, \postcode{300000}, \country{China}}}

\affil[2]{\orgdiv{Exploratory Modeling of Human-Natural Systems Research Group, Advancing Systems Analysis Program}, \orgname{ International Institute for Applied Systems Analysis (IIASA)}, \orgaddress{\city{Laxenburg}, \postcode{A-2361}, \country{Austria}}}

\affil[3]{\orgdiv{Department of Economics}, \orgname{ University of Sussex}, \orgaddress{\city{Falmer}, \postcode{BN1 9RH}, \country{UK}}}

\affil[4]{\orgdiv{Institute for Environmental Studies, Department of Spatial Economics}, \orgname{Vrije Universiteit}, \orgaddress{\city{Amsterdam}, \country{the Netherlands}}}

\affil[5]{ \orgname{Tinbergen Institute}, \orgaddress{\city{Amsterdam}, \country{the Netherlands}}}

\affil[6]{\orgname{CESifo}, \orgaddress{\city{Munich},\country{Germany}}}

\affil[7]{\orgdiv{Payne Institute for Public Policy}, \orgname{Colorado School of Mines}, \orgaddress{\city{Golden},  \state{CO}, \country{USA}}}

\affil[8]{\orgdiv{Faculty of Finance}, \orgname{City University of Macau}, \orgaddress{\street{Taipa}, \country{Macao SAR of China}}}


\abstract{Artificial intelligence (AI) interacts with climate in various ways, while a unified analytical framework of this intricate interplay is lacking. To align AI investment with climate policy, we propose such a framework integrating AI's impact on emissions, output, and climate damages into the DICE model. We distinguish between ICT-like and Industrial Revolution (IR)-like AI prospects. Calibrated to the best available evidence, we find that AI development is net polluting. Under current low abatement, ICT-like AI adds 0.1\degree C to 2100 warming, while IR-like AI adds 0.8\degree C. The associated climate costs offset roughly one-fifth and one-quarter of AI’s economic gains, respectively. Meeting the 2\degree C target saves the optimal ICT(IR)-like AI investment rate by 2100 from 3.3\% (5.1\%) under the low-abatement scenario to 3.7\% (12.7\%), indicating that mitigation is complementary to AI development. We further show that the investment trade-off between AI and abatement is driven primarily by AI’s economic prospects, not by its emissions footprint.
}




\maketitle

\begin{refsection}

\section{Introduction}\label{sec1}

\begin{figure}[h]
    \centering
    \includegraphics[width=0.98\linewidth]{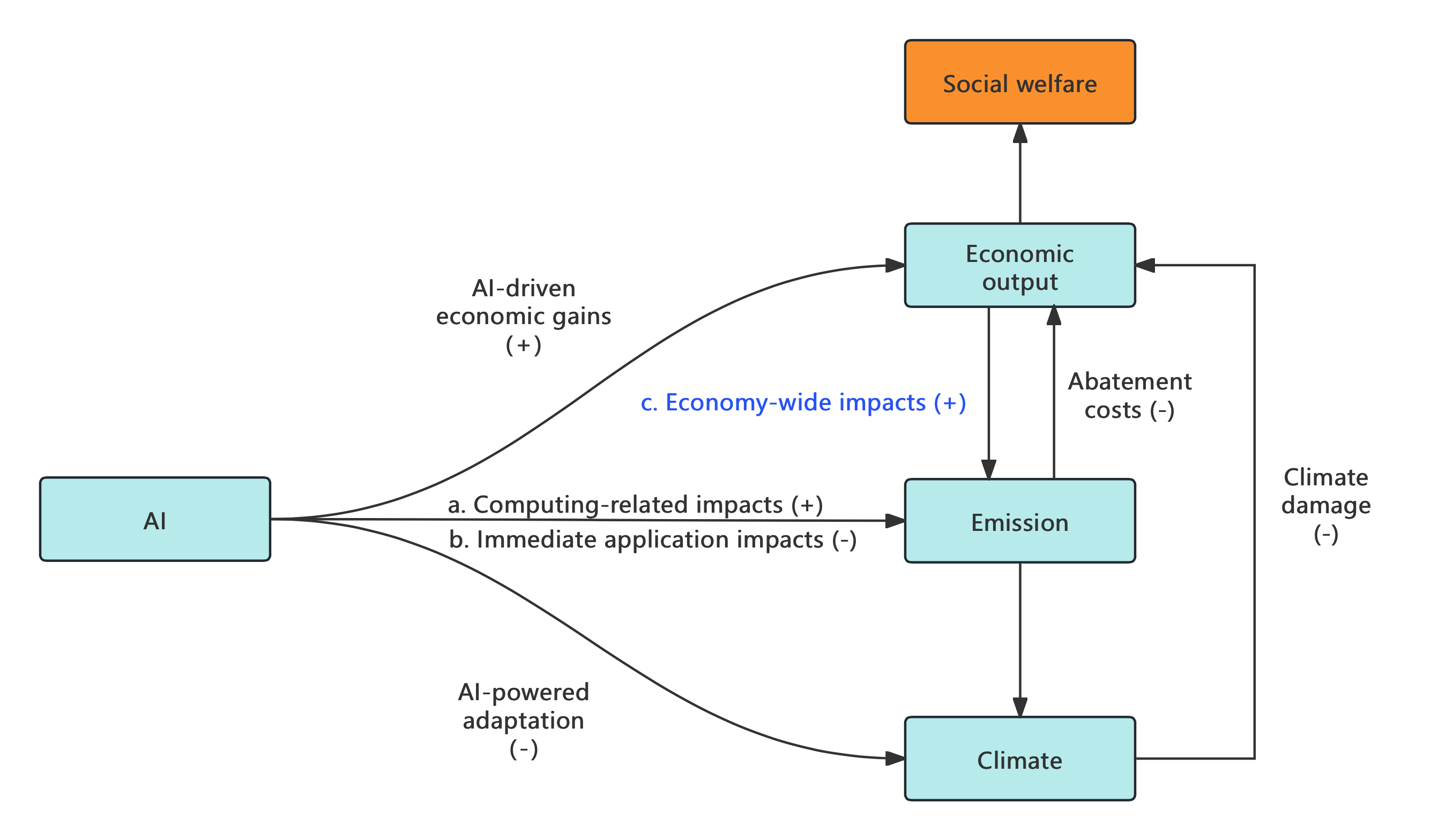}
    \caption{AI and the climate-economy system}
    \label{fig01}
\end{figure}

Artificial intelligence (AI) interacts with climate change in multiple ways. AI is energy-intensive \cite{masanet2020recalibrating, kaack2022aligning, IEA2025EnergyAI}. It directly affects both climate mitigation \cite{kaack2022aligning, rolnick2022tackling, luers2024will, IEA2025EnergyAI, stern2025green} and adaptation \cite{rolnick2022tackling, eyring2024ai}. Meanwhile, AI transforms the structure of the economy \cite{14015, acemoglu2025simple, trammell2023economic} that drives climate change, the impact of which feeds back to the economy \cite{barrage2024policies}. A unified analytical framework of this intricate interplay is lacking \cite{luers2024will}, posing a major challenge to aligning AI development with climate governance. We contribute such an integrated framework (Fig. \ref{fig01}), and apply it to a seminal integrated assessment model \cite{barrage2024policies} to derive novel policy insights.

The impact of AI on emissions is particularly complex. We follow the conceptual framework of a recent review paper \cite{kaack2022aligning} to include three categories, computing-related impact, immediate application impact, and system-level impact. The first involves GHG emissions from both energy use in machine learning and the embodied emissions in the computing infrastructure. The second relates to the short-term GHG effects of AI-related applications, with promising abatement potentials conditional on its proper use. Recent studies \cite{IEA2025EnergyAI,stern2025green,xiao2025environmental} have quantified AI-related emissions in the first two categories, despite substantial uncertainties and potentially countervailing effects. However, little is known about the system-level impact of AI, from technological development and adoption to broader lifestyle changes, which can be significant over the medium to long term \cite{luers2024will,stern2025green}. This paper not only includes the former two categories, but innovatively explores the economy-wide emission impact to represent this system-level category, because the economic system including production, technology, and policy is a critical driver of future GHG emissions \cite{nordhaus2018projections, rennert2022comprehensive}. This extended scope enables us to integrate all three categories, and provide a consistent assessment of AI's net emission impact under various economic and policy scenarios.

The complex relation between AI and climate extends beyond emissions. By empowering next-generation multiscale climate modelling \cite{eyring2024ai}, AI can improve monitoring, forecasting, and warnings of weather events (e.g. Google's GraphCast \cite{lam2023learning} \& FloodHub \cite{nearing2024global}), thereby reducing climate damage via better adaptation. In addition, AI leads to changes in economic structure \cite{trammell2023economic}, but its ultimate impact remains unclear. This deep economic uncertainty affects both mitigation and adaptation, and can further generate general equilibrium effects in the climate-economy system \cite{fernandez2024climate, NBERw33567}.

Our framework provides a unified basis for the welfare analysis of AI development and climate policy, consistent with both AI for Good \cite{ITU2025AIEnvironment} and the Sustainable Development Goals \cite{gohr2025artificial} of the United Nations. By boosting economic growth, AI increases the availability of goods and services for future consumption, this raises social welfare. However, AI development also creates emissions and affects adaptation. Therefore, balancing AI development with climate policies requires a comprehensive assessment of AI's economic benefits versus external costs\textemdash a trade-off recently recognized as one of the eight major policy challenges in the age of AI \cite{Korinek2025EconomicAI} (see Appendix \ref{secA} for a literature review).

Therefore, the paper addresses three research questions:
\begin{enumerate}
    \item \emph{What is the net impact of AI on global GHG emissions?} This requires comparing all three categories of AI's emission impact.
    \item \emph{How to align AI development with climate policies?} Specifically, if, for example, AI causes more emissions, do abatement policies curb AI investment? In turn, how does AI development shape climate policies (e.g., the social cost of carbon \cite{aldy2021keep, tol2023social, moore2024synthesis})? What drives the optimal balance?
    \item \emph{What are the climate-related welfare costs of AI development?} Do these external costs offset its economic benefits? How effective are AI-powered adaptation and mitigation policies in reducing these costs? 
\end{enumerate}

To this end, we extend the DICE-2023 model \cite{barrage2024policies} with endogenous AI growth (Fig. \ref{fig_ap02}). The inclusion of AI in an integrated assessment model is a key advance. First, we introduce AI as an intermediate production sector. To calculate GHG emissions, we differentiate the capital-embodied carbon intensity between AI and final production, which is higher in the former because AI-based data centers use more energy \cite{masanet2020recalibrating, IEA2025EnergyAI}. This explicit sectoral differentiation captures the "energy-hungry" nature of AI infrastructure. We capture the immediate application impact of AI with a reduction of the carbon intensity of the economy. The climate damage function also depends on the stock of AI capital, representing its potential to enhance societal adaptation. We consider two different channels for AI to affect economic growth \cite{14015,trammell2023economic}. One is an increase in total factor productivity growth (TFP) as AI fosters innovation \cite{babina2024artificial,labaschin2025extending}. The other automates a larger fraction of tasks in the economy \cite{nordhaus2021we,acemoglu2018race}, modelled as an increase in the capital income share. By endogenizing these channels, the model can consistently evaluate whether the productivity gains from AI are sufficient to offset its environmental costs and the potential "rebound effects" in energy consumption \cite{kaack2022aligning}.

We calibrate the model based on extensive literature surveys (see Methods and Appendix \ref{secC}). Published studies show huge disparities in AI's economic prospects\cite{aghion2024ai, acemoglu2025simple, filippucci2024miracle, korinek2023scenario}. We here consider an upper and lower bound, the latter akin to ICT (modest influence on long‑run growth, similar to ICT’s historical role) and the upper bound comparable to an Industrial Revolution (IR). Under the ICT-like pathway, annual per capita GDP growth will stabilize around 2.1\% from the second half of this century (Fig. \ref{ef04}). By comparison, the DICE model without AI projects a gradual decline to 1.8\% by 2100. Under the IR-like pathway, AI will more than double the growth rate to 4.1\% in 2100, consistent with historical industrial revolutions..

The model is kept as simple as possible, yet capturing the key interactions in the framework. We do not represent energy as an intermediate sector; instead, the energy-intensive nature of AI is reflected with a higher capital-embodied carbon intensity. This approach has the added advantage of incorporating life-cycle emissions from AI-related infrastructure, a major emission source \cite{kaack2022aligning, ITU2025AIEnvironment}. Moreover, to maintain focus on the interplay between AI and climate, we abstract from detailed AI production processes, its technology characteristics, spillover dynamics to markets, and income distribution. 

\section{Main}\label{sec2}

\subsection{Does AI cause more or less emissions?}\label{sec_warming}

\begin{figure}[htbp]
    \centering
    \begin{subfigure}[b]{0.49\textwidth}
        \centering
        \includegraphics[width=\textwidth]{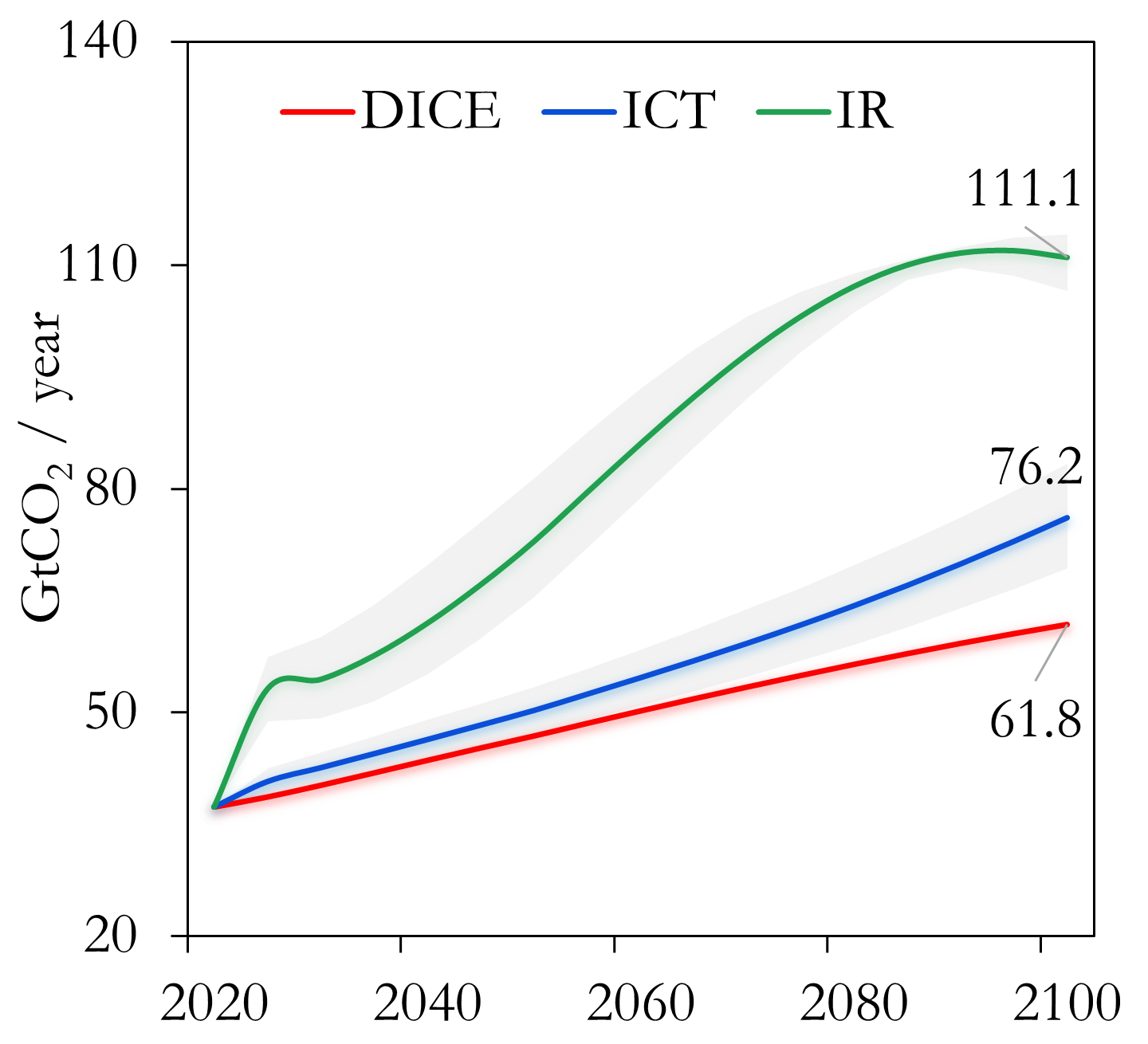}
        \caption{Annual emissions under BAU}
        \label{fig2a}
    \end{subfigure}
    \hfill
    \begin{subfigure}[b]{0.49\textwidth}
        \centering
        \includegraphics[width=\textwidth]{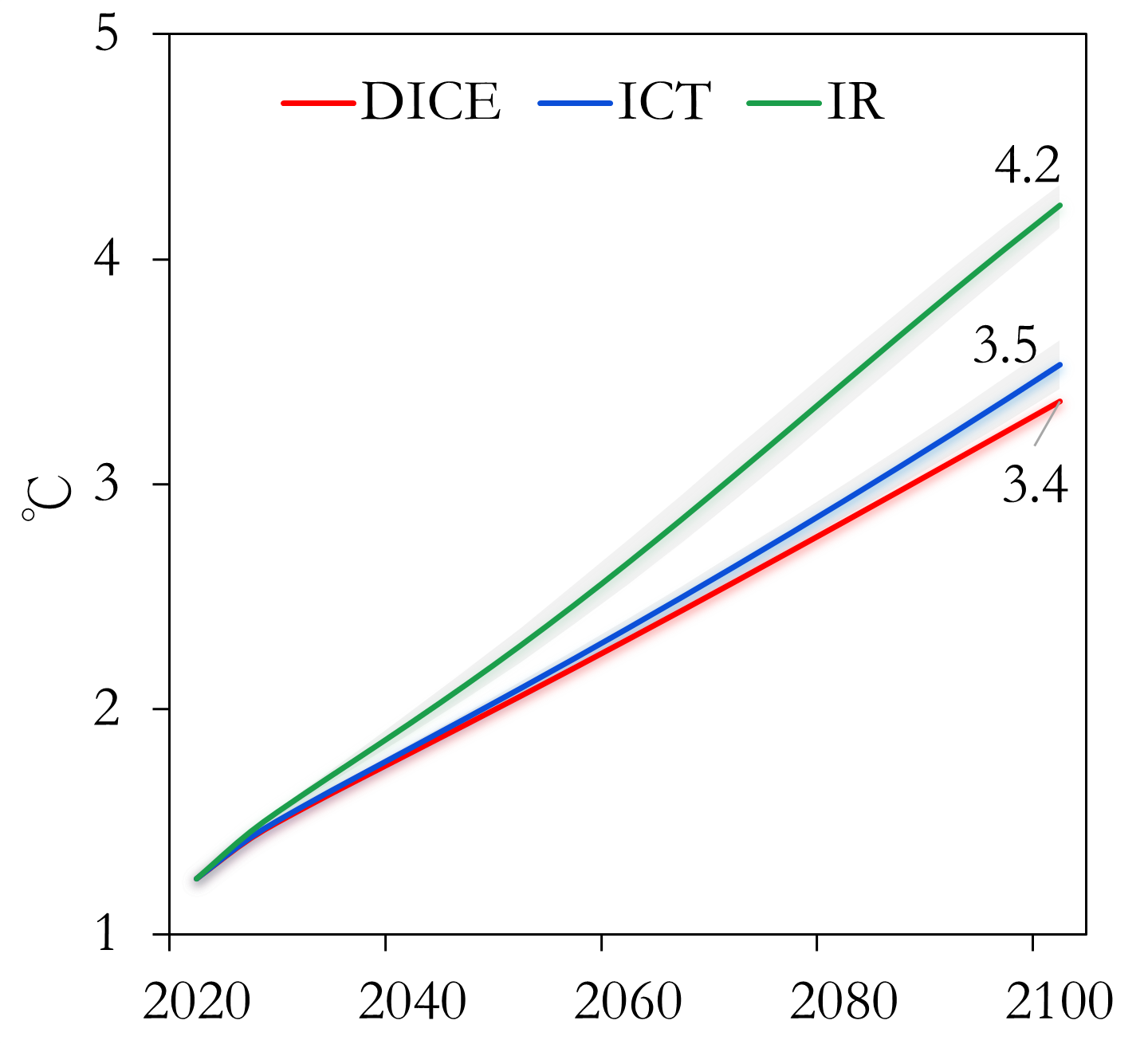}
        \caption{Global mean temperature rise under BAU}
        \label{fig2b}
    \end{subfigure}
    
    \vspace{0.1cm}

    \begin{subfigure}[b]{0.49\textwidth}
        \centering
        \includegraphics[width=\textwidth]{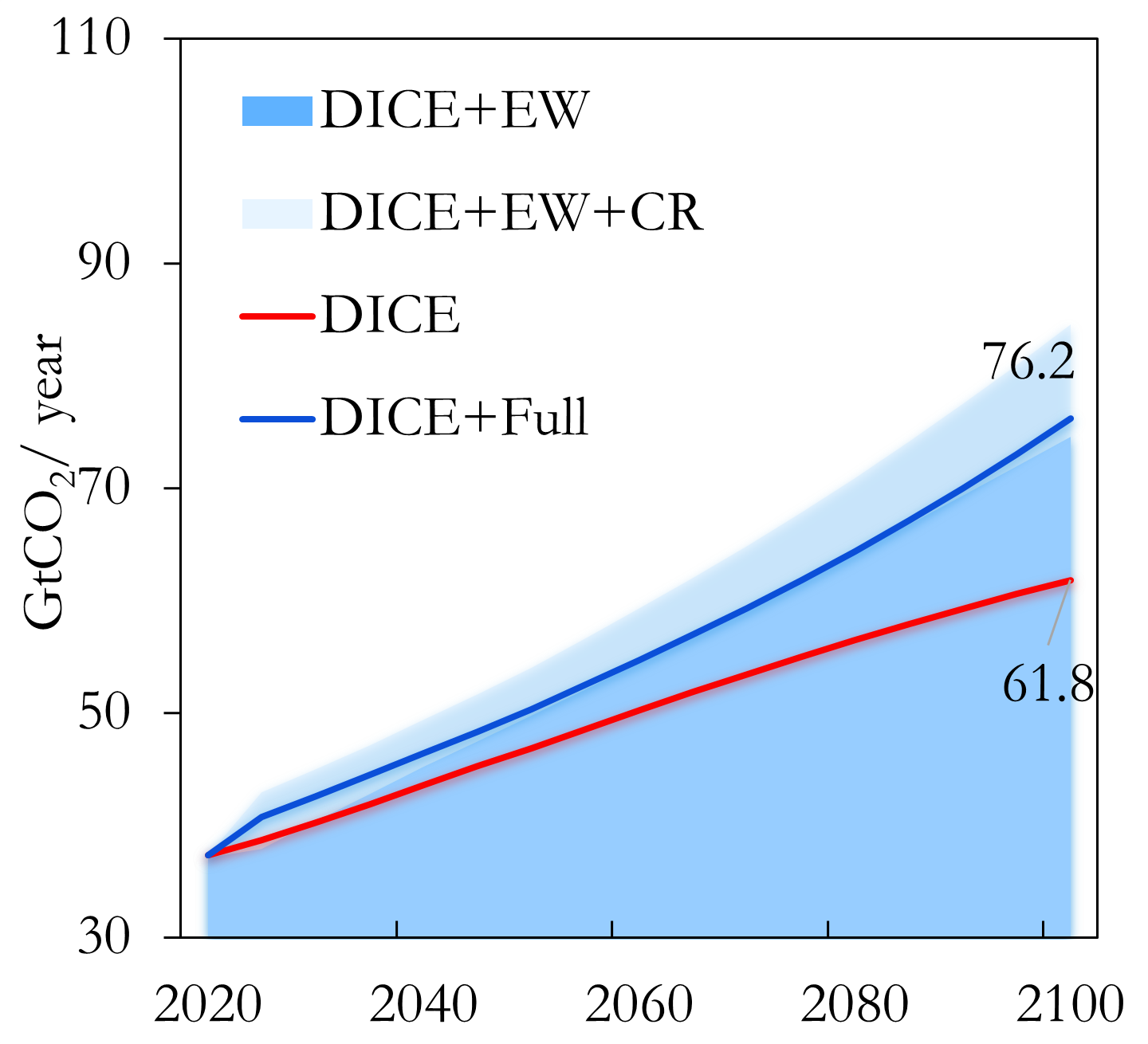}
        \caption{Emissions under the ICT-like prospect}
        \label{fig2c}
    \end{subfigure}
    \hfill
    \begin{subfigure}[b]{0.49\textwidth}
        \centering
        \includegraphics[width=\textwidth]{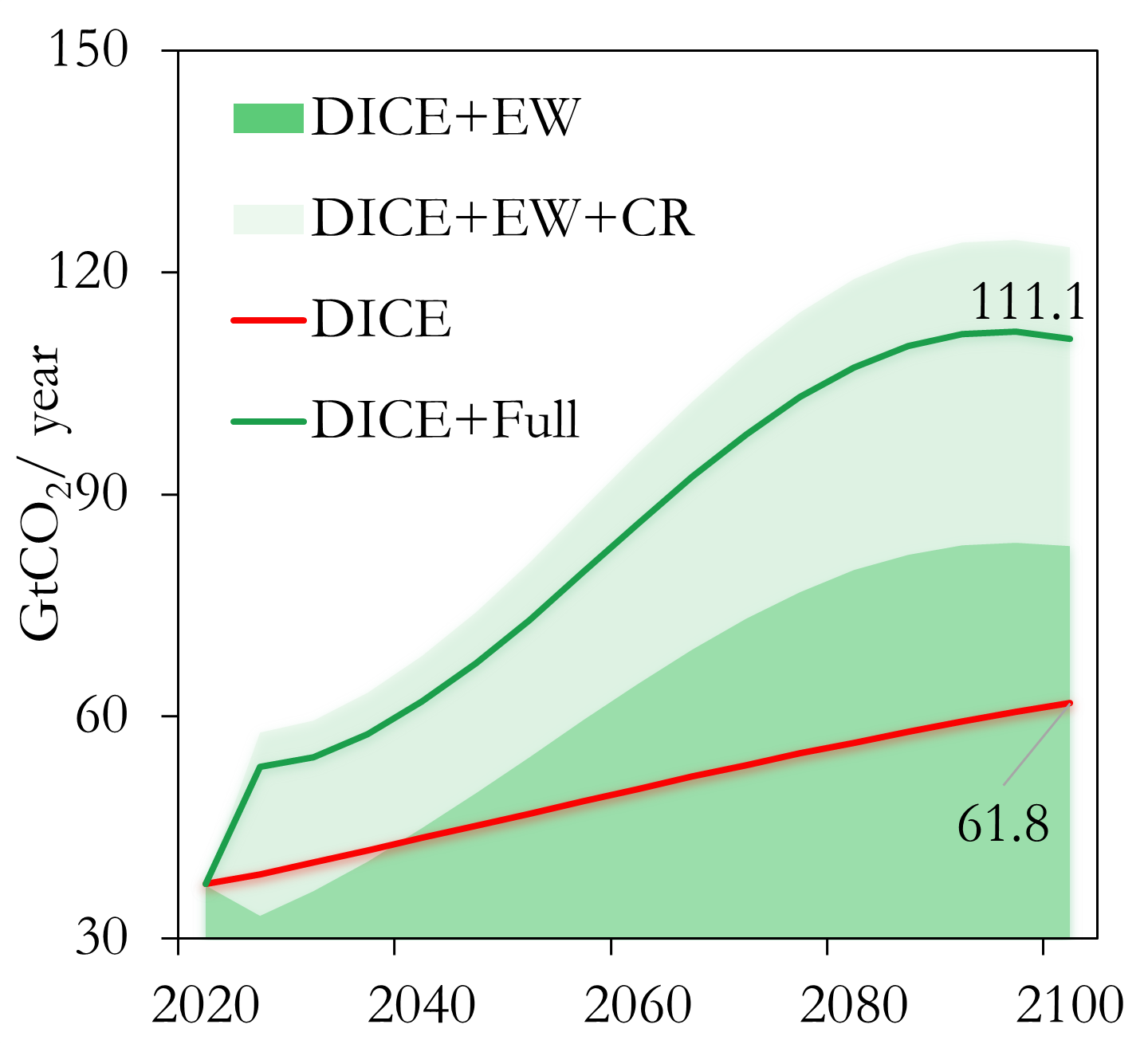}
        \caption{Emissions under the IR-like prospect}
        \label{fig2d}
    \end{subfigure}

        \caption{AI and GHG emissions}
    \label{fig2}

    \begin{tablenotes}
        \footnotesize
        \item \textbf{Note:} Annual GHG emissions and global mean temperature rise relative to the preindustrial level are calculated with the DICE-AI model, using the low abatement BAU scenario in DICE-2023 (global average carbon price of \$6/tCO\textsubscript{2} in 2020, rising at 2.5\% annually). EW and CR are short for the economy-wide impact and the computing-related impact, respectively. Their combined effects, plus the abatement from AI's immediate applications, determine the 'Full' impact. ICT projects a modest influence on long-run growth from AI, much like earlier information and communications technologies. IR posits a transformative economic impact, comparable to a new industrial revolution. The gray shaded areas in panels (a) and (b) indicate the 5th–95th percentile range of outcomes from 2,000 Monte Carlo simulations over three parameters related to AI's emissions and adaptation potential: AI-powered adaptation, the AI carbon-intensity decline rate, and AI immediate application impacts. For each parameter, we first identify the lower and upper bounds from the literature, then assume a normal distribution centered on the baseline value, and finally run the simulations based on these distributions.
    \end{tablenotes}
\end{figure}

Figure \ref{fig2} shows the projected emissions and warming impacts of AI under the two economic prospects by the DICE-AI model, assuming the BAU low abatement in DICE-2023. Relative to the DICE baseline (red line), ICT-like AI yields a higher emission path (blue line), with annual emissions in 2100 rising from 61.8 to 76.2 GtCO\textsubscript{2} (Fig. \ref{fig2a}). This raises global warming by an additional 0.1\degree C (Fig. \ref{fig2b}). In contrast, IR-like AI presents a more substantial challenge (green line). If low abatement persists, 2100 emissions would almost double to 111.1 GtCO\textsubscript{2}, further elevating the global warming to 4.2\degree C. Thus, AI will cause more emissions and hence is net polluting, without more stringent climate policies.

We also decompose the AI-associated annual emissions into distinct categories. For the ICT-like AI (Fig. \ref{fig2c}), the economy-wide (EW) emission impact of AI is substantial (darker shade), increasing by 12.8 GtCO\textsubscript{2} to 74.6 GtCO\textsubscript{2} in 2100, an increment comparable to its computing-related (CR) impact of 9.9 GtCO\textsubscript{2} (lighter shade).  Under the IR prospect (Fig. \ref{fig2d}), both the EW and CR impacts are amplified in magnitude. In initial periods, the highly profitable AI sector draws investment away from the final sector. This reallocation reduces the capital stock in the final sector, where capital depreciation outpaces new investment, generating a temporary reduction in economy-wide emissions. However, this reduction comes at the expense of surging computing-related emissions, because the emissions-intensive AI-capital is rapidly accumulating. By 2100, the EW impact increases annual emissions by 21.2 GtCO\textsubscript{2}, about half of the increase due to the CR impact (40.4 GtCO\textsubscript{2}). Together, these increases considerably diminish the relative importance of AI's abatement effects.

Our analysis highlights a critical yet overlooked role of AI's economy-wide emission impact. Increased economic output increases energy use as an essential productive factor and AI-driven automation spurs the use of energy-using machines, equipment, and robotics. Although the economy-emission nexus is well documented in both scenarios \cite{dellink2017long, o2017roads, koch2023ssp} and economics literature \cite{nordhaus2018projections}, and AI's economic potential is recognized \cite{14015, trammell2023economic}, the transmission from AI-induced economic growth to emissions remains surprisingly underexplored. 

\subsection{The interplay between AI and climate policies}

\begin{figure}[htbp]
    \centering
    \includegraphics[width=0.98 \linewidth]{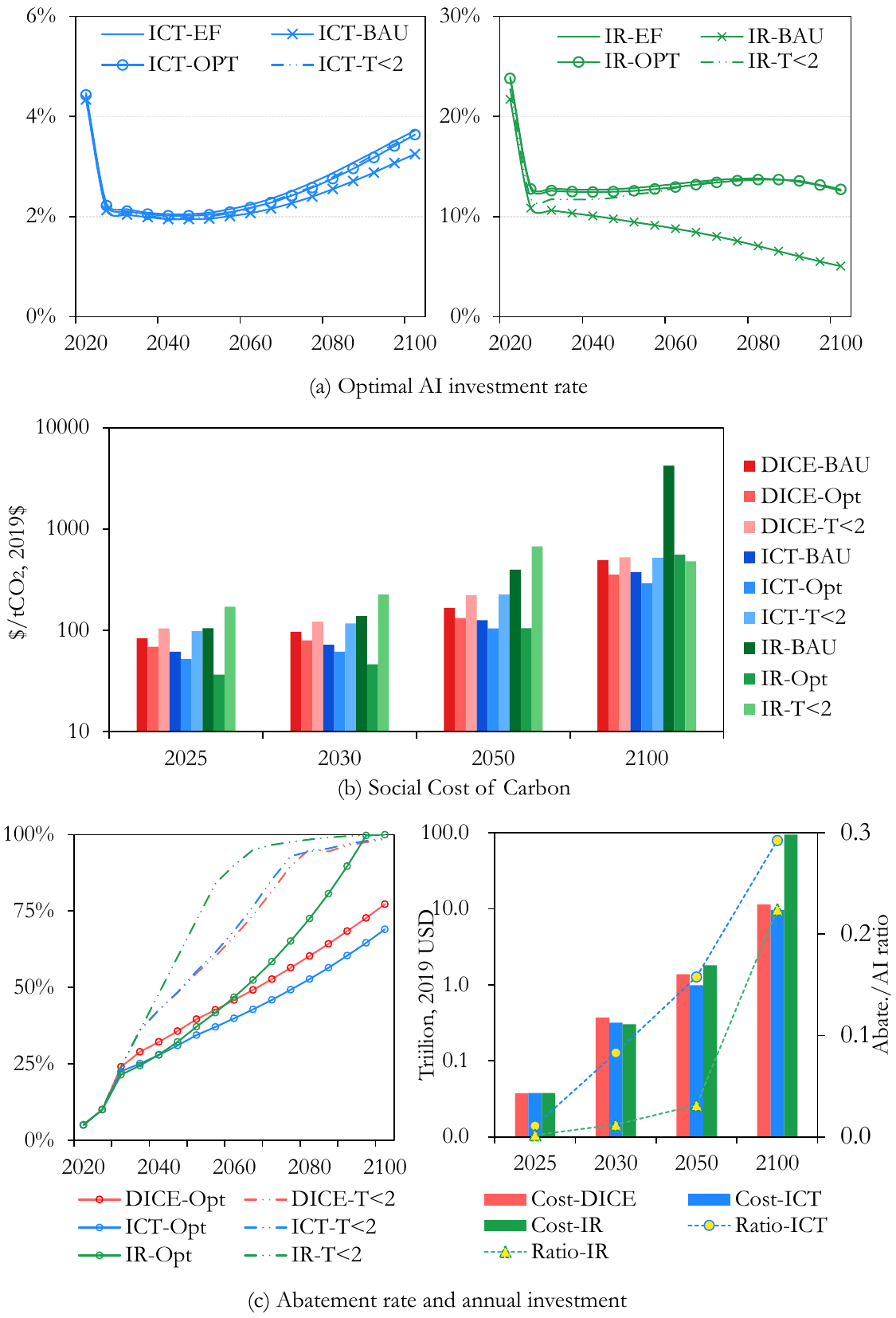}
    \caption{The interplay between AI and climate policies}
    \label{fig_3}
  
    \begin{minipage}{\textwidth}
        \footnotesize
        \textbf{Note:} ICT assumes a modest long-run growth effect of AI, similar to earlier information and communication technologies, while IR assumes a transformative effect akin to a new industrial revolution. EF denotes the economic frontier without climate damages and serves as the pure growth benchmark. BAU denotes the low-abatement business-as-usual scenario in DICE-2023, Opt the welfare-maximizing carbon price, and $T<2$\degree C the 2\degree C temperature target. Panel (a) shows the optimal AI investment rate as a share of final output in DICE-AI. Panel (b) reports the social cost of carbon (SCC) in DICE and DICE-AI under different policy scenarios. The left-hand figure in panel (c) shows the abatement rate, while the right-hand figure reports annual abatement costs under Opt for DICE, ICT, and IR; the dotted line shows carbon mitigation investment per dollar of AI investment for ICT and IR. 

    \end{minipage}
    
\end{figure}

AI development and climate policy interact in two ways. We begin by analyzing the impact of climate policy on AI. Given our prior findings that AI is net polluting, one might expect that GHG mitigation policies hinder AI investment. The results in Figure \ref{fig_3} reveal the opposite.

We simulate the optimal AI investment rate (AI investment as a share of final output) under different scenarios using the DICE-AI model. For the ICT-like prospect (left, Panel (a)), the economic frontier (EF) scenario (no climate impact, solid blue line) yields a U-shaped AI investment rate, due to the combined effects of accelerated TFP growth and capital deepening (see Appendix \ref{secAI_inv}). However, when climate impacts are included under a business-as-usual (BAU) low-abatement policy, the AI investment rate shows a marked drop\textemdash by 0.12 pp to 1.96\% in 2050, and by a further 0.47 pp to 3.25\% in 2100. Notably, whether it be an optimal carbon tax or a 2\degree C warming constraint, these emission-reducing climate policies largely reverse the decline in AI investment observed under BAU, resulting in AI investment rates of 3.64\% by 2050 and 3.66\% by 2100, respectively. A similar pattern is observed under the IR-like prospect (right), where the decline under BAU is more acute relative to EF, and is similarly reversed by mitigation policies. Thus, climate policies are in effect complementary to AI development. In the BAU, accelerating global warming inflicts climate damages, undermining the future return on AI capital. Effective climate policies, by curbing damages, preserve the future profitability of AI investment.


We next examine how AI development affects climate policies. Panel (b) reports the social cost of carbon (SCC) as a yardstick of climate policy \cite{aldy2021keep, kelleher2025social}. It measures the consumption loss due to an additional tonne of carbon emissions \cite{barrage2024policies}. Under welfare-maximizing conditions, the SCC coincides numerically with the optimal carbon tax, the Pigou tax; in other scenarios, the SCC is the damage done by emitting an additional tonne of carbon dioxide.


Under BAU with the ICT prospect, the SCC is lower than DICE (i.e., 61 v.s. 84 \$/tCO\textsubscript{2} in 2025, 126 vs. 166 \$/tCO\textsubscript{2} in 2050, and 376 vs. 492 \$/tCO\textsubscript{2} in 2100). This reduction occurs because AI-powered adaptation outweighs the emission-intensifying impact of ICT-like modest economic growth. Without such adaptation, climate damages are amplified, due to the curvature of the damage function implying increasing marginal damages at higher temperatures, resulting in a higher SCC path (e.g., 83, 171, and 531 in 2025, 2050, and 2100, respectively; see Table \ref{tab1} in 'Extended results'). Reduced climate vulnerability resulting from AI‑powered adaptation also lowers the optimal carbon tax under ICT‑like AI. The optimal carbon tax is 105 \$/tCO\textsubscript{2} in 2050 and 291 \$/tCO\textsubscript{2} in 2100, lower than the corresponding DICE values of 132 and 356 \$/tCO\textsubscript{2}. By comparison, differences are smaller between DICE and ICT-like AI under a 2\degree C warming constraint, because cost-effective policies are less sensitive to climate vulnerability.

IR-like AI markedly alters the SCC’s time profile and magnitude. Substantially larger damages under BAU drive the SCC upward, because the GHG emissions generated by IR‑like explosive growth are so extensive that their climate impact overwhelms any adaptive benefits from AI. Compared to DICE, the SCC more than doubles to 425 \$/tCO\textsubscript{2} in 2050 and soars to 4107 \$/tCO\textsubscript{2} in 2100. This large SCC increase under BAU does not proportionally raise the optimal carbon tax, though the impact remains substantial. Under optimal climate policy, IR-like AI lowers the SCC in the near term and raises it over the mid‑ to long term. By 2100, the optimal carbon tax (557 \$/tCO\textsubscript{2}) exceeds the DICE value (356 \$/tCO\textsubscript{2}) by half. When a 2\degree C warming cap is imposed, the SCC is higher in early decades but lower by 2100, due to DICE's backstop technology. Because IR‑like AI stimulates explosive economic growth, it expands the base of GHG emissions, necessitating a higher emission control rate that will activate the backstop sooner. Such a constraint acts as a kink in the damage function, elevating the SCC \cite{barrage2024policies}. 



The abatement rate confirms our results for the SCC (Panel (c)). For the ICT-like AI,  a more lenient pathway results with AI-enhanced adaptation, whereas a stricter abatement path is obtained without considering adaptation. By comparison, IR-like AI produces a more stringent abatement path from 2060 onward. By 2100, while the DICE model still permits an optimal residual emission level of 23\%, IR-like AI requires net-zero emissions. Under the 2\degree C warming target, AI development accelerates the timeline for deep decarbonization (see Fig. \ref{ef2} in 'Extended results'). 



While AI-powered adaptation has considerable potential to alleviate damages, it does not necessarily justify a more lenient climate policy. Persistent uncertainties around the discount rate, damage estimates, whether AI will materialize as ICT or IR, the impact of AI-powered adaptation, and the interactions among them, can still support more stringent abatement policies than the DICE benchmark. However, when AI's economic impact is very promising, it exacerbates the mitigation pressure, particularly under a 2\degree C warming target.


We further examine how climate mitigation investment responds to AI development. We consider the amount of carbon mitigation investment per dollar directed to AI investment. Intuitively, one might expect this ratio to be lower for ICT-like AI than for IR-like AI, because the latter represents a large emission base. Counterintuitively, Panel (c) displays the opposite pattern (e.g., 0.01 v.s. 0.002 in 2025, 0.16 v.s. 0.03 in 2050, and 0.29 v.s. 0.22 in 2100). Indeed, the ratio more meaningfully reflects which type of investment yields higher marginal returns. Given that IR-like AI heralds greater revenue from AI investment, a rational social planner invests more heavily in AI, thereby lowering \emph{the mitigation‑investment‑per‑dollar-AI-investment ratio}. Nevertheless, in absolute terms, annual abatement investment remains higher under IR-like AI, consistent with the larger emission base associated with stronger economic growth. 


 

\subsection{Welfare analysis of AI development}


Using the permanent-consumption-equivalent (PEV) method \cite{barrage2020optimal}, we perform an intertemporal welfare comparison. 


\begin{figure}[htbp]
    \centering
    \begin{subfigure}[b]{0.49\textwidth}
        \centering
        \includegraphics[width=\textwidth]{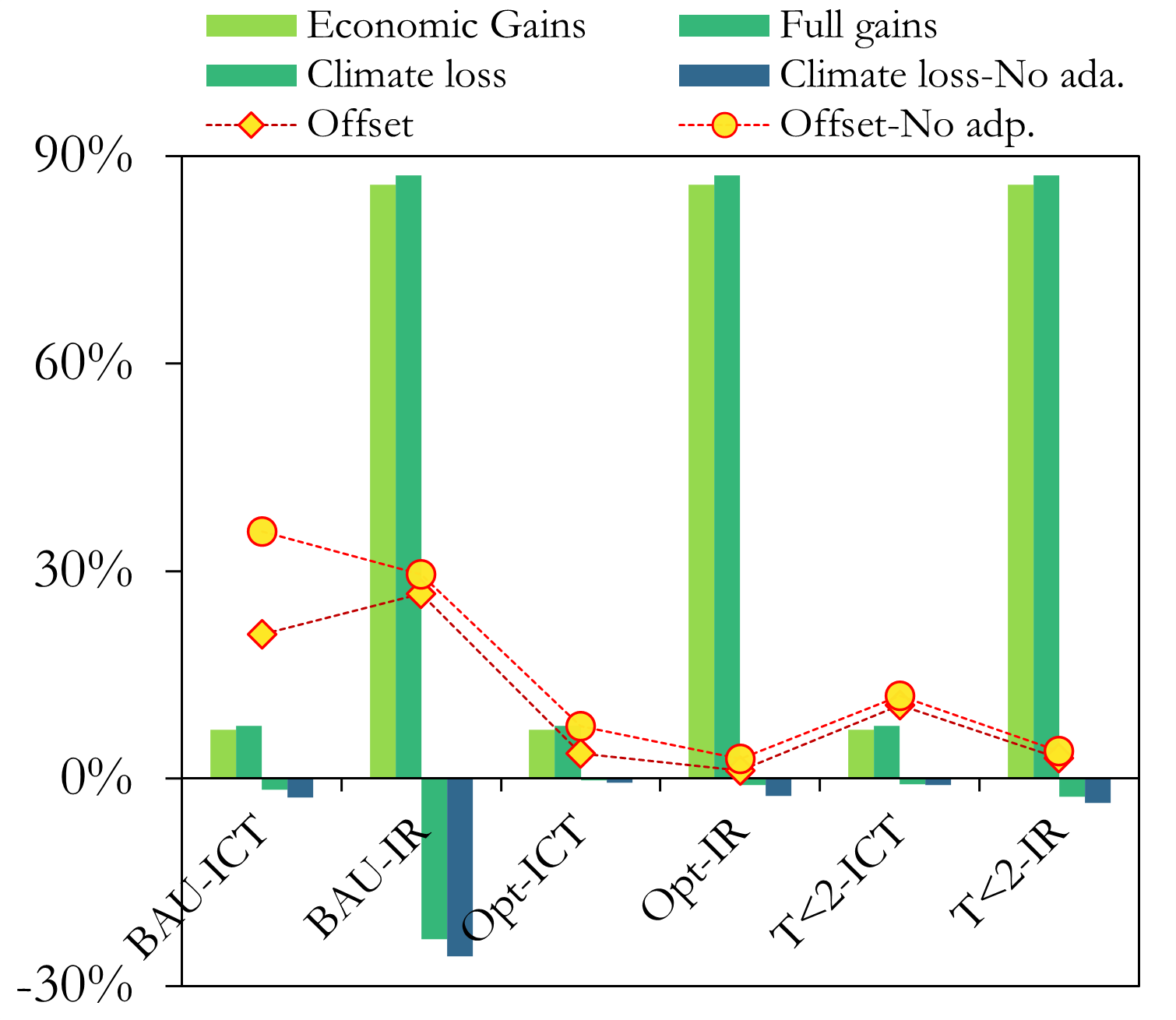}
        \caption{Welfare}
        \label{fig4a}
    \end{subfigure}
    \hfill
    \begin{subfigure}[b]{0.49\textwidth}
        \centering
        \includegraphics[width=\textwidth]{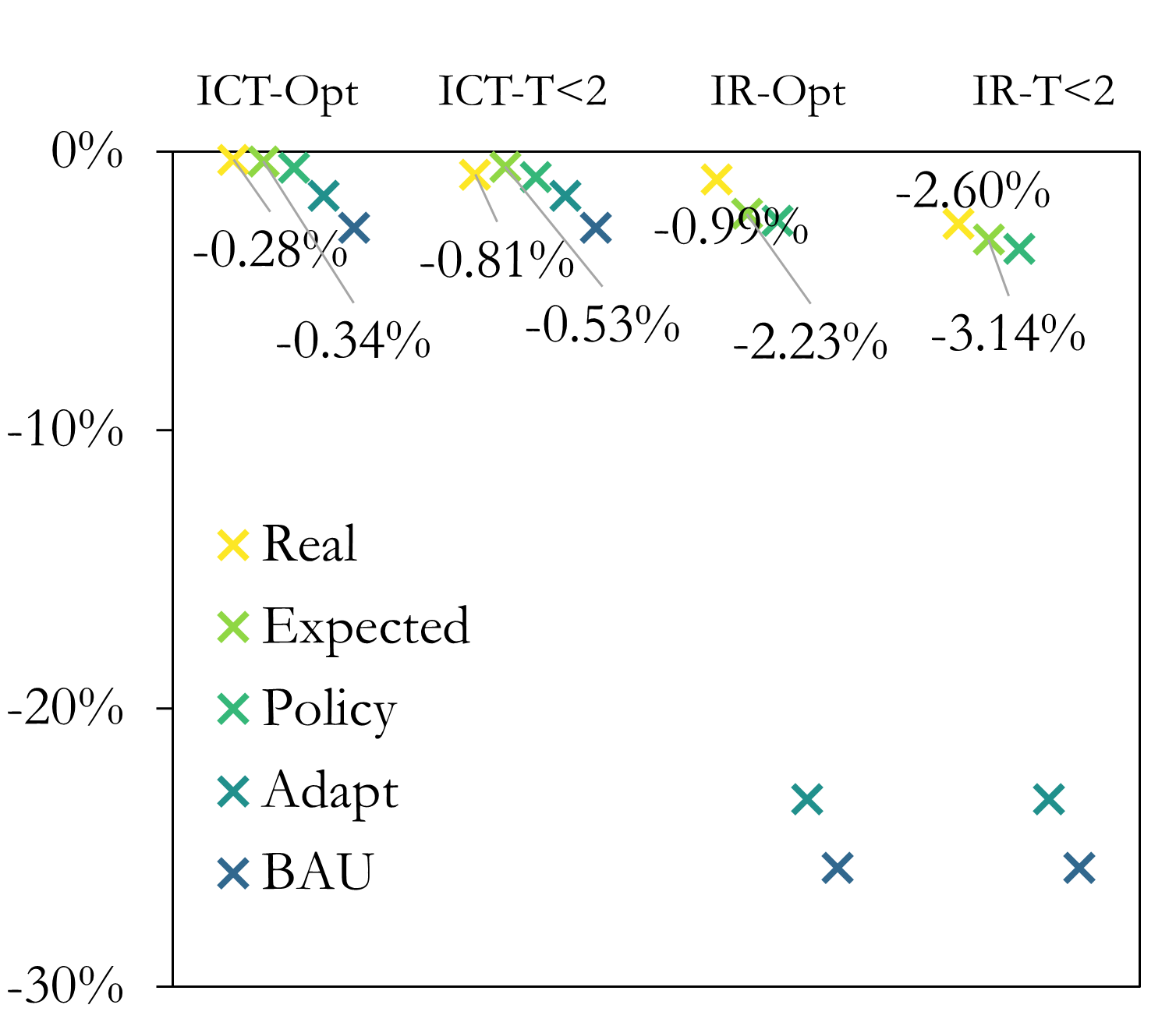}
        \caption{Policy \& adaptation}
        \label{fig4b}
    \end{subfigure}
    
    \caption{Climate-related welfare loss under different AI development prospects}
    \label{fig4}
    
    \vspace{2mm}
    \begin{minipage}{\textwidth}
        \footnotesize
        \textbf{Note:} ICT projects a modest influence on long-run growth from AI, much like earlier information and communications technologies. IR assumes a transformative economic impact, triggering a new industrial revolution. BAU represents the current business-as-usual low abatement policy as implemented in DICE-2023. OPT employs an optimal carbon price to maximize social welfare. T$<$2 caps the global mean temperature rise at 2\degree C above its preindustrial level. Using the permanent-equivalent-variation method following Ref. \cite{barrage2020optimal}, the figure displays the climate-related welfare loss from AI development, expressed relative to the consumption path of the DICE-2023 optimal scenario. The loss captures both climate damages from AI-induced warming and associated abatement costs (see 'Methods'). 'Policy' excludes AI-powered adaptation, whereas 'AI adapt' assumes the BAU policy. The 'Expected' losses from combining climate policy and AI adaptation are calculated by multiplying their individual residual loss shares, while the 'Real' loss denotes the modeled outcome under their combined implementation.
    \end{minipage}
\end{figure}

Our baseline DICE-AI model reveals that under BAU, the climate welfare loss attributable to polluting AI is substantial, even when AI’s adaptation potential is fully utilized (Fig. \ref{fig4} and Table  \ref{et01}). By aggravating global warming further, GHG emissions from ICT-like and IR-like AI would induce a permanent consumption loss of 2.73\% and 25.73\%, respectively, if AI-powered adaptation is excluded. When adaptation is enabled, consumption loss remains considerable at 1.59\% and 23.24\%. These results reflect the combined effect of AI-driven warming and the convexity of the climate damage function, where damages increase progressively with temperature. Thus, the climate loss for IR-like AI is an order of magnitude greater than that for ICT-like AI.


Nevertheless, well-designed climate policy can mitigate these losses and, combined with AI-powered adaptation, produce multiplier effects. An optimal carbon tax alone reduces the permanent consumption loss to 0.58\% and 2.47\%, respectively. The expected consumption loss from combining abatement policy and AI adaptation\textemdash calculated by multiplying their individual residual loss shares\textemdash is 0.34\% for ICT-like and 2.23\% for IR-like AI. However, the modeled outcome combining both shows even lower realized losses of 0.28\% and 0.99\%. Thus, combining abatement policy with AI adaptation offers a promising future for sharply reducing climate welfare losses.  

We further quantify how climate-related costs offset the economic gains from AI development, highlighting the critical role of climate policy in safeguarding AI-driven welfare benefits. Although AI adaptation reduces climate damages, residual climate-related losses still offset a notable share of AI economic gains. Under BAU, climate costs due to AI-induced warming offset approximately one-fifth (ICT) or one-quarter (IR) of AI economic gains. In contrast, climate policy preserves most of these gains, reducing the offset to 3.64\% (ICT) and 1.14\% (IR) under the optimal tax, and to 10.60\% and 2.98\% under the 2\degree C warming limit. 

\subsection{Sensitivity}
Despite its abatement potential, AI is a net GHG source. A thought experiment (Fig. \ref{ef03}) asks how powerful AI’s immediate application must be to offset its other two emission categories under BAU. Reversing this net-polluting verdict proves exceptionally difficult, and likely unrealistic for an IR-like future. This requires ICT-like and IR-like AI to reduce aggregate carbon intensity by around 28\% and 90\%, respectively. This is equivalent to global carbon intensity in 2050 falling from today's level in Cambodia to that of Greece under ICT, or from Pakistan's to Sweden's under IR. Our model assumes no extra abatement costs; historically, an economy with effective carbon taxes like Sweden took 4 and 26 years for comparable declines.

\begin{figure}[htbp]
    \centering
    \includegraphics[width=0.85 \linewidth]{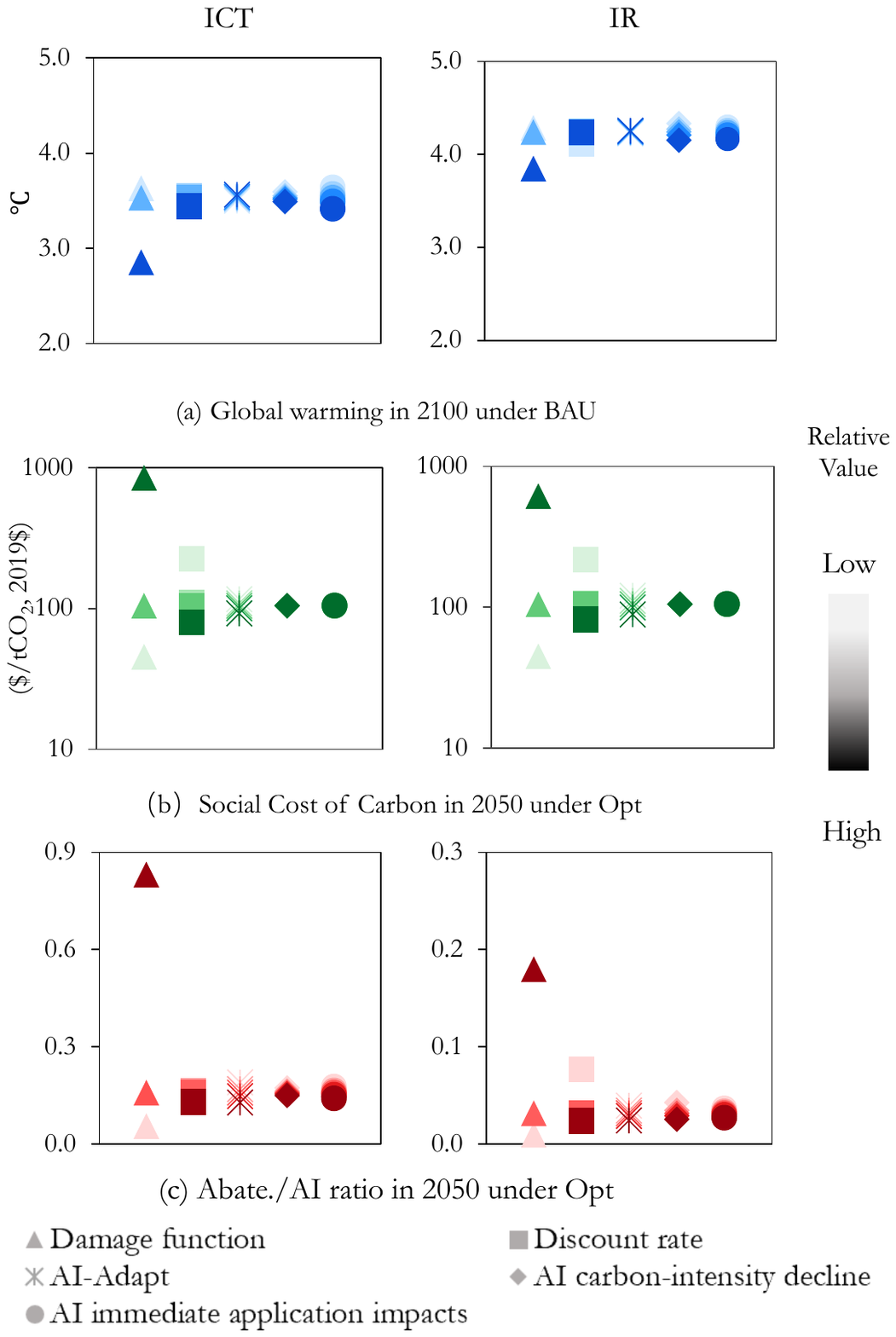}
    \caption{Sensitivity Analysis}
    \label{fig_5}
  
    \begin{minipage}{\textwidth}
        \footnotesize
        \textbf{Note:} ICT projects a modest influence on long-run growth from AI, much like earlier information and communications technologies. IR posits a transformative economic impact sufficient to trigger a new industrial revolution. The darker the mark, the higher the value it represents. For the damage function, we consider three values in the sensitivity analysis: our baseline specification and two alternatives from \cite{Tol2024ClimateMeta}. For the discount rate, we consider four values: our baseline value (the mean recommended by economists, \cite{drupp2018discounting}), the mean recommended by philosophers, \cite{nesje2023philosophers}, and the values used by \cite{stern2007economics} and \cite{nordhaus2018projections}. For AI-powered adaptation, the AI carbon-intensity decline rate, we first determine the lower and upper bounds of each parameter based on literature review, then assume a normal distribution with the baseline value as the mean, and finally select the values at the 2.5th, 25th, 50th, 75th, and 97.5th percentiles for the sensitivity analysis. Further details on these parameters are provided in the Calibration section (see Appendix \ref{sensi_ex}).

    \end{minipage}
    
\end{figure}

We further consider five parameters: damage-output elasticity, discount rate, AI adaptation, AI carbon intensity decline, and AI’s immediate applications (Fig. \ref{fig_5}). Damage-output elasticity consistently dominates all outcome metrics under both ICT‑like and IR‑like scenarios. Through a feedback loop, higher elasticity reduces capital accumulation and emissions, moderating warming under low abatement.

For temperature, with the ICT‑like prospect, the uncertainty of the immediate application impact dominates that of AI carbon intensity; this pattern is reversed in the IR‑like case due to massive capital accumulation. The optimal SCC is mainly driven by the discount rate and damage elasticity, not AI emission externalities. \emph{The mitigation‑investment‑per‑dollar-AI-investment ratio} is robust across AI‑related uncertainties. Thus, AI‑specific emission and adaptation uncertainties do not fundamentally shift optimal carbon policy. More profoundly, the broader development pathway (ICT‑like vs. IR‑like) represents a deeper source of uncertainty for warming trajectories and the AI‑investment‑mitigation balance.

\section{Discussion and Conclusion}\label{sec12}


The more optimistic the economic prospects of AI development, the greater the resulting emissions, intensifying the mitigation pressure. AI-powered adaptation alleviates but does not reverse this.

Well-designed climate policy is complementary to AI development. By curbing climate change and its damages, such policy preserves the future economic value generated by AI investments. To fully harness AI's economic benefits, robust climate policy is essential; either an optimal carbon tax or 2\degree C warming limit reduces AI's climate-related welfare costs. The primary factor in the trade-off between climate action and AI investment is not AI's own emissions impact, but the relative marginal return on investment in decarbonization versus AI capital.

This study develops an integrated framework and implements it through the DICE-AI model to understand the interactions between AI and climate. This parsimonious, integrated model focuses on macro-level climate effects and policy feedbacks, while inevitably abstracting from the full complexity of AI's real-world impacts. It, however, provides a flexible starting point for systemic analysis. We invite the empirical and modeling communities to build upon this framework to refine and elaborate on the specific channels through which AI will shape—and be shaped by—our climate future.

\section{Methods}\label{sec11}
Based on the DICE-2023 model \cite{barrage2024policies}, we apply the proposed framework to build the DICE-AI model. The model adopts a standard optimal growth framework, in which a representative household’s intertemporal welfare is defined over an infinite path of per capita consumption. A social planner is posited to maximize this intertemporal welfare objective. The original DICE model has only one final production sector. Here we introduce an intermediate AI production sector that combines capital $K_{AI,t}$ and labor $L_{AI,t}$ using standard Cobb-Douglas technology for production:
\begin{equation}
   y^{AI}_t=A_{AI,t} K_{AI,t}^\alpha L_{AI,t}^{1-\alpha} 
\end{equation}
where $A_{AI,t}$ denotes the technology level. To restrict our attention to the interplay between AI and climate, we do not delve into the details of AI production. However, this simplification suffices to accommodate three main perspectives in our framework, as explained below. 

First, we introduce how AI affects total economic production. To maintain the closest comparability with DICE, we preserve the Cobb-Douglas production technology so that the final production is given by:
\begin{equation}
    Y_t=A_t(y_t^{AI}) \times K_t^{\gamma(y_t^{AI})}  L_t^{1-\gamma(y_t^{AI})}
\end{equation}
Existing studies \cite{14015,trammell2023economic} suggest two principal channels through which AI reshapes economic growth, by enhancing total factor productivity (TFP) growth and fostering automation in tasks. AI-driven TFP growth can be alternatively interpreted as
the automation of research and development. Motivated by the observation that
ideas—and hence growth—become harder to find over time \cite{bloom2020ideas}, we model AI-induced TFP growth as, 
\begin{equation}
    \frac{A_{t+1}-A_t}{A_t}=g_{base,t}+\theta_{TFP}\times (y_t^{AI} -y_t^{AI,base}) \times A_t^{-\beta}
\end{equation}
where $g_{base,t}$ is the exogenously-given TFP growth forecasts from DICE-2023. Given $\beta>0$, the term $A_t^{-\beta}$ captures the economic intuition that better ideas are increasingly harder to find \cite{bloom2020ideas}. Thus, more AI output is required to sustain higher TFP growth. For better calibration, we infer the implicit level of AI output in the DICE-2023 model, denoted by $y_t^{AI,base}$. Thus, AI's net economic impact depends on the difference between the AI output level endogenously determined in equilibrium $y_t^{AI}$ and this exogenously-given level $y_t^{AI,base}$. 

Besides, because AI development is expected to facilitate automation, we model that:
\begin{equation}
    \gamma(y_t^{AI})=\gamma_{asymp}-\gamma_{adj}\times e^{-\theta_{\gamma}\times (y_t^{AI} -y_t^{AI,base})}
\end{equation}
such that a growing AI sector asymptotically increases the capital income share by $\gamma_{adj}$ to $\gamma_{asymp}$. The flip side of the coin is the declining labor income share, as manual labor will gradually be replaced by machinery and robotics \cite{14015,acemoglu2018race,acemoglu2025simple}. 

Second, we conceive of AI as a stock adaptation \cite{felgenhauer2014modeling}. Following the DICE-2023 model, the damage function measures a proportional loss of final output, but is now given by:
\begin{equation}
    D(T_t,K_{AI,t})=\left[(1-a_3)+  a_3 \left( \frac{K_{AI,0}}{K_{AI,t}}\right)^{a_4} \right] a_1T_t^{a_2}. 
\end{equation}This implies that as AI capital accumulates, it will asymptotically alleviate a fraction of $a_3\in(0,1)$ in climate damage. Without adaptation, climate damage is governed by $a_1T_t^{a_2}$, as in DICE.

Third, we introduce AI's emission impact. To calculate aggregate greenhouse gas emission, the original DICE model associates emissions with total economic production. By comparison, DICE-AI leverages the concept of capital-embodied carbon intensity \cite{seto2016carbon,tong2019committed,ye2023allocating}, which is different between the final sector and the AI sector. In addition, to capture the immediate application impact, we model that the AI output will lower the carbon intensity of the overall economy without incurring additional abatement costs. Thus, total carbon emissions are 
\begin{equation}
    E_t^{Ind}=(1-\Omega(y_t^{AI}))\times (\sigma^K_t \times K_t + \sigma^{K_{AI}}_t \times K_{AI,t})
\end{equation}
where $\sigma^i_t$ is the capital-embodied carbon intensity in sector $i$. Because the AI sector consumes more energy and also emits more throughout its life-cycle, one expects $\sigma^K_t<\sigma^{K_{AI}}_t$. $\Omega(y_t^{AI})$ gauges the fraction of reduced emissions. More specifically, we use:
\begin{equation}
     (1-\Omega(y_t^{AI}))= \psi_{asymp}+\psi_{adj}\times e^ {-\theta_{sav} \times (y_t^{AI} -y_t^{AI,base})}
\end{equation}
such that a fraction of $\psi_{adj}$ emissions will be abated when the AI output increases infinitely. $\theta_{sav}$ is the speed scale of the immediate application effect. The economy-wide emission impact is naturally included, once we allow AI to affect economic production.

Last, we explain how capital and labor are allocated or accumulated. For capital, we follow the putty-clay growth model \cite{atkeson1999models,baldwin2020build} that capital accumulation is individual and separate between sectors, which gives:
\begin{align}
    K_{t+1}&=(1-\delta)K_t+I_t\\
    K_{AI,t+1}&=(1-\delta)K_{AI,t}+I_t^{AI}
\end{align}
The economic intuition behind is that once economic output is used for investment either in final or AI sectors, it builds up stranded assets, which, combined with capital-embodied carbon intensity, leads to irreversibility effects on emissions \cite{baldwin2020build,van2020stranded}. For simplicity, we assume a constant ratio of labor force in AI sector $L_{AI,t}=\nu L_t^{Total}$ and:
\begin{equation}
    L_{AI,t} + L_t =L_t^{Total}
\end{equation}
In each period, total economic production net of climate damages and abatement costs is allocated between consumption, general investment, and AI investment:
\begin{equation}
    Y_t \times (1-D(T_t,K_{AI,t})-\Lambda(\mu_t))=C_t+I_t+I_t^{AI}
\end{equation}
As in DICE-2023, $\Lambda(\mu_t)$ is the abatement cost depending on the emission abatement rate $\mu_t$. Thus, an amount of $E_t=(1-\mu_t)E_t^{Ind}$ unabated emissions will flow into the atmosphere, causing the global mean temperature rise:
\begin{equation}
    T_{t}=\Phi\left(E_0,E_1,...,E_t\right)
\end{equation}
Carbon and climate dynamics are adopted from the DICE-2023 model directly, with the equilibrium climate sensitivity of 3\degree C and transient climate response of 1.8\degree C.

 The latest DICE vintage \cite{barrage2024policies} utilizes a time-varying path of discount rates, which affects future investment incentives in an intertemporal optimization framework. To isolate the specific impact of AI from such dynamic discounting effects, we adopt a constant discount rate in our analysis. The benchmark value for this rate is taken from the economist survey by \cite{drupp2018discounting}. For sensitivity analysis, we also consider alternative rates recommended by philosophers \cite{nesje2023philosophers}, the DICE-2016 vintage \cite{nordhaus2018projections}, and Stern \cite{stern2007economics}. The initial capital stock in AI sector is calculated using perpetual inventory method, applied to AI investment time series data from Stanford University Human-Centered Artificial Intelligence (HAI) database \cite{AIIndex2025}. 
We conducted a comprehensive review of the literature on data-center carbon emissions  (Table \ref{tab_cal_AI_em}). Reported estimates for around 2020 range from 38 to 330 MtCO\textsubscript{2}, primarily due to differences in system boundaries and the treatment of embodied emissions. Considering the global scope of our analysis and our inclusion of embodied emissions, we calibrate the AI sector's initial emissions to 220 MtCO\textsubscript{2}. These two values pin down the initial carbon intensity of AI capital, which is 0.2316 kg/\$ (2019 PPP). AI sector is assumed to have the same exogenous labor-augmented productivity as the final sector.

 To ensure comparability with DICE, we initially seek to mimic the results of the DICE-2023 model with our DICE-AI model. By fixing the AI investment rate in DICE-AI at its initial observed level \cite{AIIndex2025}, we recalibrate the TFP and its growth, as well as the carbon intensity of general capital, so that total economic output and aggregate emissions from 2020 to 2100 match between the two models. Then, we relax the constraint on AI investment rate to calibrate AI's economic impact in equilibrium, by adjusting the TFP growth and the capital income share. In addition, together with the AI-capital emission intensity obtained above, the difference between the DICE's total emissions and AI sector's emissions residually determines the general capital emissions intensity.
 
 Existing projections of AI‑driven growth tend to be conservative, often extrapolating from the historical performance of ICT. Nevertheless, the future remains highly uncertain. The rapid diffusion of generative AI in recent years raises the prospect that AI could trigger an Industrial Revolution–like transformation. For the ICT-like prospect, the parameters governing AI’s economic impact are calibrated based on a comprehensive literature review (Table \ref{tab_cal_eco}), drawing on estimates from prominent AI economists \cite{aghion2024ai,acemoglu2025simple,JonesTonetti_2025_Automation,Korinek2025EconomicAI}. The chosen values align with published estimates of productivity gains and capital‑share shifts. For the IR scenario, direct empirical evidence is scarce, so we adopt a historical‑analogue approach, calibrating parameters based on evidence from the First, Second, and Third Industrial Revolutions. The resulting parameter values generate growth trajectories comparable in magnitude to those observed in previous industrial revolutions (Figure \ref{ef04b}).  Consequently, for the ICT-like prospect, global GDP per capita growth (before climate damages) stabilises at roughly 2.1\% per annum over this century. By comparison, the IR prospect projects average annual growth rising to 4.1\%, more than double the DICE baseline of 1.8\% . 

Both the computing-related and the economy-wide emission impacts of AI are determined upon the calibration of its economic prospects and capital-embodied carbon intensity. Now we proceed to calibrate AI's abatement potentials. Based on a survey of the literature (Table \ref{tab_cal_AI_abate}), the immediate-application impacts of AI are set such that its adoption reduces 2035 global emissions by 3 GtCO\textsubscript{2} in equilibrium under BAU scenario. This 3 GtCO\textsubscript{2} reduction lies midway between the estimates reported by Stern et al. (2025) \cite{stern2025green} and the IEA (2025) \cite{IEA2025EnergyAI}. For clarity, we note three distinct channels of decarbonization in the DICE-AI model. The first is the exogenously-given decline in the carbon intensity embodied in both AI capital and general capital over time (Fig. \ref{ef06}). The second is the AI-driven abatement described above, which is achieved without additional cost and acts as a positive externality of AI output. The third is conventional costly abatement, represented by the abatement cost function and determined as an endogenous variable in the model.
 
To calibrate AI's adaptation benefits, we combine recent sectoral damage estimates from Ref. \cite{tol2023social} with the aggregate damage function of DICE to estimate the maximum potential of AI to reduce climate impacts across sectors (Table \ref{tab_cal_AI_adapt}). This results in an estimated reduction of aggregate climate damage from 3.12\% to 1.81\% of output at 3°C of warming. We assume that the AI capital stock in 2050 will realize half of this maximum adaptation potential in equilibrium, which pins down two damage function parameters, $a_3=0.42$ and $a_4=0.1736$. For sensitivity analysis, we draw on two representative damage functions from the updated meta-analysis \cite{howard2025methodology} and apply the same adaptation parameters. Additional details and parameters of the DICE-AI model are provided in Appendix \ref{secC}.

The DICE‑AI model retains the same mathematical structure as the standard DICE framework and constitutes a concave dynamic optimization problem over the calibrated parameter space. The model is implemented in GAMS and solved using several large‑scale nonlinear programming solvers, all of which converge to identical optimal trajectories, confirming the robustness and uniqueness of the solution.

 Our analysis further relies on detailed post-processing of the DICE-AI model outputs to enable welfare comparisons across scenarios. This is accomplished in two stages. First, we employ the Permanent-Equivalent Variation (PEV) method following Ref. \cite{barrage2020optimal} to compare welfare between any two scenarios with distinct consumption paths, $\{C_t^{Base}\}_{t=0}^\infty$ and $\{C_t^{*}\}_{t=0}^\infty$. This method requires adopting a unified definition of the social welfare:
 \begin{equation}
      W\triangleq\sum_{t=0}^{\infty} \beta^t U(C_{t}) L_t
 \end{equation}
 It solves for a constant scaling factor $\varphi$ that equates the welfare of the compared path to a proportionally reduced version of the baseline path: $W^{*}=\sum_{t=0}^{\infty} \beta^t U(C_{t}^{*}) L_t=\sum_{t=0}^{\infty} \beta^t U\left((1-\varphi)C^{Base}_{t}\right) L_t$. Here, $\varphi$ gauges the permanent consumption loss of path $\{C_t^{*}\}_{t=0}^\infty$ relative to $\{C_t^{Base}\}_{t=0}^\infty$. The PEV method provides an internally consistent metric for comparing welfare under vastly different assumptions about growth, damages, and policy. While the consumption-equivalent SCC offers a simpler alternative for direct welfare calculation, it is a marginal concept. Its use can introduce significant approximation inaccuracies in aggregate welfare assessment when emission differences between scenarios are large.

Second, we implement a five-step procedure to structure the comparison across different scenarios:
 
 \textbf{Step 1} Run the DICE-AI model under the optimal carbon price. This represents a world incorporating AI's full economic, emission, and adaptation impacts. The resulting AI-sector investment rates are saved.
 
 \textbf{Step 2} Run the DICE-AI model with a fixed temperature pathway from standard DICE and the predetermined AI investment rates from Step 1. This constructs an 'ideal' counterfactual world where AI delivers its economic and adaptation benefits without exacerbating climate change. The resulting consumption path is saved.
 
 \textbf{Step 3} The welfare difference between the consumption path from this ideal world (Step 2) and that from the standard DICE optimal scenario (without AI) defines the maximum potential welfare gain from AI. This gain combines AI's direct economic contribution and its adaptation benefits. The adaptation component can be isolated by simply deactivating the adaptation function within the DICE-AI model.
 
 \textbf{Step 4} Compare the consumption path from any policy scenario featuring AI's full impact to the ideal-world path from Step 2. The resulting loss in welfare measures the climate‑related costs attributable to AI.  Under the optimal carbon tax policy scenario, these costs reflect only the climate damages from AI-induced warming. Under the 2\degree C warming limit scenario, the costs incorporate both damages and the required abatement expenditures.
 
 \textbf{Step 5} Normalize all computed welfare gains and climate-related costs using the consumption path from the standard DICE optimal scenario as the numeraire, ensuring consistent scaling.

\backmatter

\bmhead{Acknowledgements}

We are grateful to Angus C. Chu, Reyer Gerlagh, Xinyue Hao, Markus Leibrecht, Le Li, Carles Mañó-Cabello, Karl W. Steininger, Ulrich Wagner,  Wenli Xu, and Zaifu Yang and all participants at the 14th Mannheim Conference on Energy and the Environment, the 7th World Congress of Environmental and Resource Economists (WCERE 2026), the UM Econ-CityU FOF joint workshop for constructive comments and feedback. Huiying Ye appreciates the funding from National Natural Science Foundation of China (No.72504082). The views expressed are our own and do not reflect the views of the supporting agencies or authors' affiliations.

\printbibliography[title=References]

\end{refsection}

\bigskip

\newpage
\begin{appendices}

\begin{refsection}

\section{Extended results}\label{secD}

\subsection{Economic prospects of AI} \label{apendixa1}
\begin{figure}[htbp]
    \centering
    \begin{subfigure}[b]{0.49\textwidth}
        \centering
        \includegraphics[width=\textwidth]{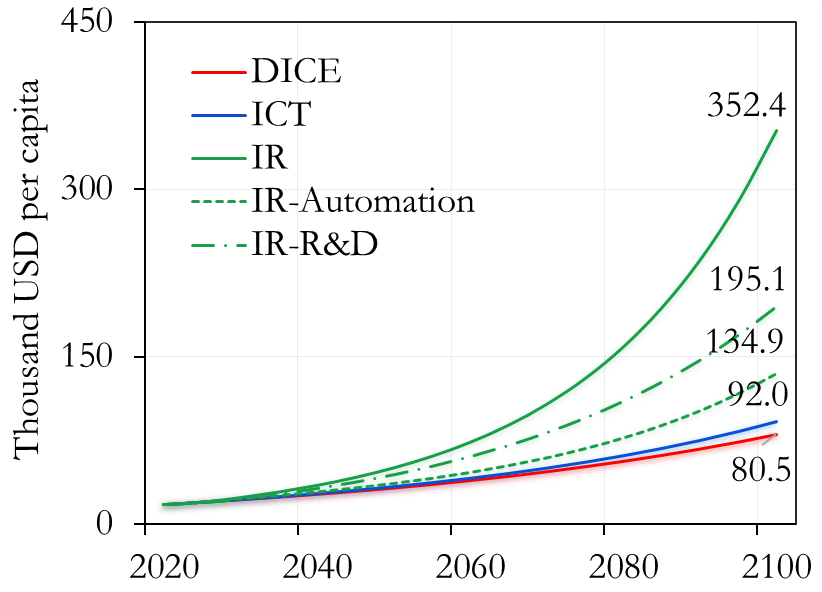}
        \caption{Y/L}
        \label{ef04a}
    \end{subfigure}
    \hfill
    \begin{subfigure}[b]{0.49\textwidth}
        \centering
        \includegraphics[width=\textwidth]{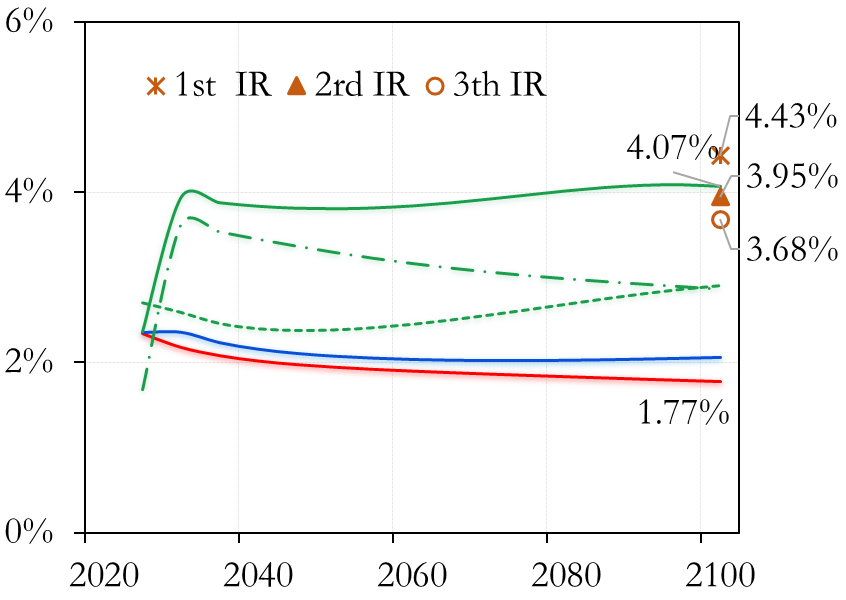}
        \caption{Growth of Y/L}
        \label{ef04b}
    \end{subfigure}

    \caption{Economic prospects under ICT and IR}
    \label{ef04}

    \begin{minipage}{\textwidth}
        \footnotesize
        \textbf{Note:} ICT projects a modest influence on long-run growth from AI, much like earlier information and communications technologies. IR posits a transformative economic impact sufficient to trigger a new industrial revolution. 'Automation' only accounts for AI's impact on increasing capital income share, while 'R\&D' merely includes AI's impact on accelerating TFP growth. Three markers corresponding to past industrial revolutions in Panel (b) are obtained by multiplying the respective amplification factors (2.5, 2.24, and 2.08) by the 2100 per capita GDP growth rate in DICE (1.8\%).
    \end{minipage}
    
\end{figure}

\subsection{Social cost of carbon}
\begin{table}[htbp]
  \centering
  \caption{Social cost of carbon (\$/tCO\textsubscript{2}, 2019\$)}
    \begin{tabular}{cclcccccc}
    \toprule
    \multicolumn{3}{c}{Scenario} & 2025  & 2030  & 2050  & 2100  & $T_{max}^{21st}$ (\degree C) & $T_{max}$ (\degree C) \\
    \midrule
    \multirow{3}[2]{*}{DICE} & \multicolumn{2}{c}{BAU} & 84    & 97    & 166   & 492   & 3.4  & 5.4 \\
          & \multicolumn{2}{c}{Opt} & 69    & 79    & 132   & 356   & 2.6  & 2.7 \\
          & \multicolumn{2}{c}{$T<2$\degree C } & 105   & 122   & 222   & 524   & 2.0  & 2.0 \\
    \midrule
    \multirow{9}[6]{*}{ICT} & \multirow{2}[2]{*}{BAU} & No adaptation & 83    & 97    & 171   & 531   & 3.5  & 6.1 \\
          &       & Baseline & 61    & 72    & 126   & 376   & 3.5  & 6.4 \\
\cmidrule{2-9}          & \multirow{6}[2]{*}{Opt} & Baseline & 53    & 61    & 105   & 291   & 2.8  & 3.0 \\
          &       & Philosopher & 55    & 65    & 110   & 301   & 2.8  & 3.0 \\
          &       & Stern & 118   & 139   & 225   & 463   & 2.4  & 2.4 \\
          &       & Nordhaus  & 40    & 46    & 80    & 238   & 2.9  & 3.3 \\
          &       & Alt damage 1 & 29& 33& 54& 148& 3.2& 4.2\\
          &       & Alt damage 2 & 599& 631& 849& 1796& 1.8& 2.5\\
\cmidrule{2-9}          & $T<2$\degree C   & Baseline & 98    & 117   & 226   & 521   & 2.0  & 2.0 \\
    \midrule
    \multirow{9}[6]{*}{IR} & \multirow{2}[2]{*}{BAU} & No adaptation & 118   & 153   & 425   & 4107  & 4.2  & 6.9 \\
          &       & Baseline & 105   & 138   & 396   & 4239  & 4.2  & 6.9 \\
\cmidrule{2-9}          & \multirow{6}[2]{*}{Opt} & Baseline & 36    & 46    & 105   & 557   & 3.3  & 3.3 \\
          &       & Philosopher & 37    & 46    & 106   & 558   & 3.3 & 3.3 \\
          &       & Stern & 76    & 98    & 217   & 874   & 2.6  & 2.6 \\
          &       & Nordhaus  & 29    & 36    & 82    & 457   & 3.5  & 3.5 \\
          &       & Alt damage 1 & 19
& 24& 50& 274& 4.5& 4.8\\
          &       & Alt damage 2 & 313& 360& 616& 2287& 2.0& 2.0\\
\cmidrule{2-9}          & $T<2$\degree C   & Baseline & 171   & 227   & 677   & 481   & 2.0  & 2.0 \\
    \bottomrule
    \end{tabular}%
  \label{tab1}%

\begin{tablenotes}
        \footnotesize
        \item \textbf{Note:} ICT projects a modest influence on long-run growth from AI, much like earlier information and communications technologies. IR posits a transformative economic impact, triggering a new industrial revolution. BAU represents the current business-as-usual low abatement policy as implemented in DICE-2023. OPT employs an optimal carbon price to maximize social welfare. $T<2$\degree C caps the global mean temperature rise at 2\degree C above the preindustrial level. The 'No adaptation' scenario precludes AI-powered adaptation from reducing climate damages, as opposed to the Baseline scenario, which assumes that AI development fulfills half of its adaptation potential by 2050. Our baseline simulation adopts the discount rate parameters based on a survey of economists \cite{drupp2018discounting}. For sensitivity tests, we also consider discounting parameters advocated by philosophers \cite{nesje2023philosophers}, used in the Stern Report \cite{stern2007economics} and in earlier work by Nordhaus \cite{nordhaus2018projections}. Additionally, we include two alternative damage estimates from a recent updated meta-analysis of climate damages \cite{Tol2024ClimateMeta}. Here, $T_{max}^{21st}$ denotes the highest temperature rise within this century, while $T_{max}$ refers to the maximum temperature rise projected beyond this century.
    \end{tablenotes}  
\end{table}%
\newpage

\subsection{AI investment rate}\label{secAI_inv}

\begin{figure}[htbp]
    \centering
    \begin{subfigure}[b]{0.49\textwidth}
        \centering
        \includegraphics[width=\textwidth]{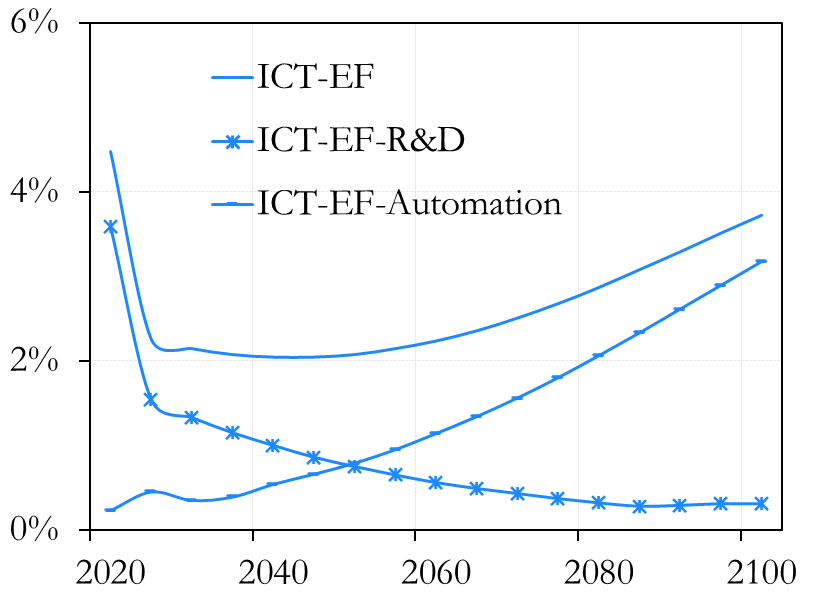}
        \caption{ICT-like prospect}
        \label{ef1a}
    \end{subfigure}
    \hfill
    \begin{subfigure}[b]{0.49\textwidth}
        \centering
        \includegraphics[width=\textwidth]{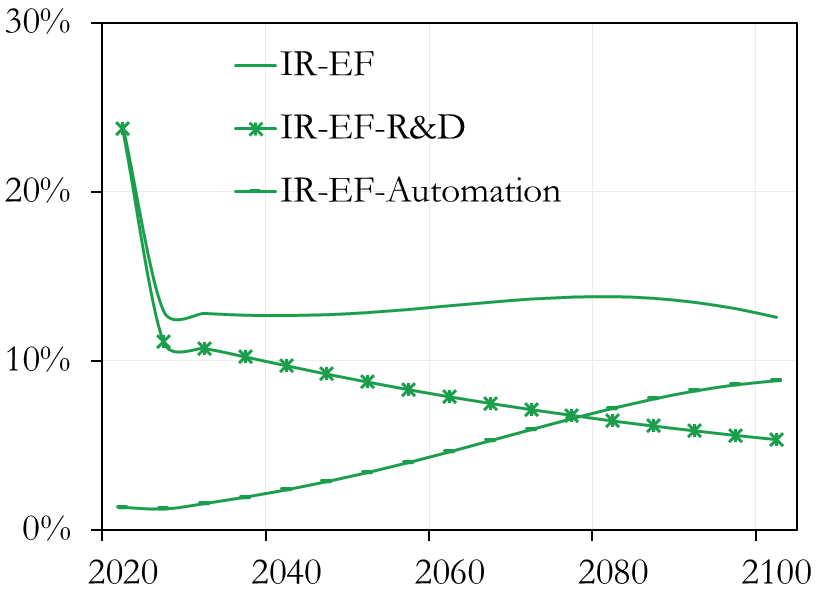}
        \caption{IR-like prospect}
        \label{ef1b}
    \end{subfigure}
    
    \caption{AI investment rate in the economic frontier}
    \label{ef1}

    \vspace{2mm}
    \begin{minipage}{\textwidth}
        \footnotesize
        \textbf{Note:} The figure presents the resulting optimal AI investment rate (AI investment as a ratio of final output) in the economic frontier of the DICE-AI model. ICT projects a modest influence on long-run growth from AI, much like earlier information and communications technologies. IR posits a transformative economic impact sufficient to trigger a new industrial revolution. 'Automation' only accounts for AI's impact on increasing capital income share, while 'R\&D' merely includes AI's impact on accelerating TFP growth.
    \end{minipage}
    
\end{figure}

The curve of the optimal AI investment rate resulting from the DICE-AI model is shaped by two distinct forces (Fig. \ref{ef1}). First, AI can accelerate total factor productivity (TFP) growth by enhancing innovation. If AI only raises TFP growth, the optimal investment rate declines over time because the marginal return to R\&D diminishes as ideas become harder to find \cite{bloom2020ideas}, though early-stage incentives remain strong. Second, AI development can automate an increasing fraction of tasks across the economy, gradually raising the capital income share. If AI only drives capital deepening, the optimal investment rate generally increases over time, as the full economic effect of automation unfolds gradually, particularly in the mid- to long-term. Combining these two forces yields an optimal AI investment path that first decreases and later rises in the long run.

\newpage
\subsection{Abatement rate}

\begin{sidewaystable}[htbp]
  \caption{Abatement rate and annual investment}
    \begin{tabular*}{\textheight}{@{\extracolsep\fill}clcccccccccccc}
    \toprule
    \multicolumn{2}{c}{\multirow{2}[4]{*}{Scenario}} & \multicolumn{4}{c}{DICE}      & \multicolumn{4}{c}{ICT}       & \multicolumn{4}{c}{IR} \\
\cmidrule{3-6}\cmidrule{7-10}\cmidrule{11-14}    \multicolumn{2}{c}{} & 2025  & 2030  & 2050  & 2100  & 2025  & 2030  & 2050  & 2100  & 2025  & 2030  & 2050  & 2100 \\
\midrule
    \multirow{5}[2]{*}{Control rate} & Opt.-No adapt. & 10\%  & 24\% & 40\%  & 77\%  & 10\%  & 24\%  & 40\%  & 79\%  & 10\%  & 24\%  & 45\%  & 100\% \\
          & Opt.  & \textbackslash{} & \textbackslash{} & \textbackslash{} & \textbackslash{} & 10\%  & 23\%  & 34\%  & 69\%  & 10\%  & 21\%  & 37\%  & 100\% \\
          & Opt-Alt damage 1 & 10\%  & 18\%  & 27\%  & 53\% & 10\%  & 15\%  & 23\%  & 46\%  & 10\%  & 14\%  & 24\%  & 72\% \\
          & Opt-Alt damage 2 & 10\%  & 24\%  & 72\%  & 100\% & 10\%  & 24\%  & 72\%  & 100\% & 10\%  & 24\%  &72\%  & 100\% \\
          & $T<2$\degree C   & 10\%  & 24\%  & 55\%  & 98\%  & 10\%  & 24\%  & 55\%  & 99\%  & 10\%  & 24\%  & 72\% & 100\% \\
    \midrule
    \multirow{5}[2]{*}{\makecell{Annual investment\\(Trillion 2019\$)}} & Opt.-No adapt. & 0.04  & 0.37  & 1.38  & 11.40  & 0.04  & 0.38  & 1.43  & 13.74  & 0.04  & 0.41  & 2.91  & 93.68  \\
          & Opt.  & \textbackslash{} & \textbackslash{} & \textbackslash{} & \textbackslash{} & 0.04  & 0.32  & 0.99  & 9.70  & 0.04  & 0.30  & 1.81  & 94.75  \\
          & Opt-Alt damage 1 & 0.04  & 0.18  & 0.53  & 4.38  & 0.04  & 0.12  & 0.35  & 3.35  & 0.04  & 0.10  & 0.56  & 40.36 \\
          & Opt-Alt damage 2 & 0.04  & 0.32  & 5.41  & 15.64  & 0.04  & 0.35  & 6.44  & 23.74  & 0.04  & 0.39  & 9.72  & 80.60  \\
          & $T<2$\degree C   & 0.04  & 0.37  & 3.16  & 21.56  & 0.04  & 0.37  & 3.38  & 24.76  & 0.04  & 0.40  & 9.78  & 93.32  \\
\cmidrule{2-14}    \multirow{5}[2]{*}{Abate./AI ratio} & Opt.-No adapt. & \textbackslash{} & \textbackslash{} & \textbackslash{} & \textbackslash{} & 0.01  & 0.10  & 0.24  & 0.42  & 0.002  & 0.02  & 0.05  & 0.23  \\
          & Opt   & \textbackslash{} & \textbackslash{} & \textbackslash{} & \textbackslash{} & 0.01  & 0.08  & 0.16  & 0.29  & 0.002  & 0.01  & 0.03  & 0.22  \\
          & Opt-Alt damage 1 & \textbackslash{} & \textbackslash{} & \textbackslash{} & \textbackslash{} & 0.01  & 0.03  & 0.05  & 0.10  & 0.002  & 0.00  & 0.01  & 0.09  \\
          & Opt-Alt damage 2 & \textbackslash{} & \textbackslash{} & \textbackslash{} & \textbackslash{} & 0.01  & 0.08  & 0.83  & 0.85  & 0.002  & 0.02  & 0.18  & 0.26  \\
          & $T<2$\degree C   & \textbackslash{} & \textbackslash{} & \textbackslash{} & \textbackslash{} & 0.01  & 0.10  & 0.56  & 0.74  & 0.002  & 0.02  & 0.18  & 0.22  \\
    \bottomrule
    \end{tabular*}%
  \label{tab2}%

\begin{tablenotes}
        \footnotesize
        \item \textbf{Note:} ICT projects a modest influence on long-run growth from AI, much like earlier information and communications technologies. IR posits a transformative economic impact, triggering a new industrial revolution. BAU represents the current business-as-usual low abatement policy as implemented in DICE-2023. OPT employs an optimal carbon price to maximize social welfare. $T<2$\degree C caps the global mean temperature rise at 2\degree C above the preindustrial level. The 'No adaptation' scenario excludes AI-powered adaptation from reducing climate damages, as opposed to the Baseline scenario, which assumes that AI development fulfills half of its adaptation potential by 2050. Additionally, we include two alternative damage estimates from a recent updated meta-analysis of climate damages \cite{Tol2024ClimateMeta}. The 'Abate./AI ratio' measures the amount of carbon mitigation investment per dollar directed to AI investment.
    \end{tablenotes}
\end{sidewaystable}

\begin{figure}[htbp]
    \centering
    \begin{subfigure}[b]{0.4\textwidth}
        \centering
        \includegraphics[width=\textwidth]{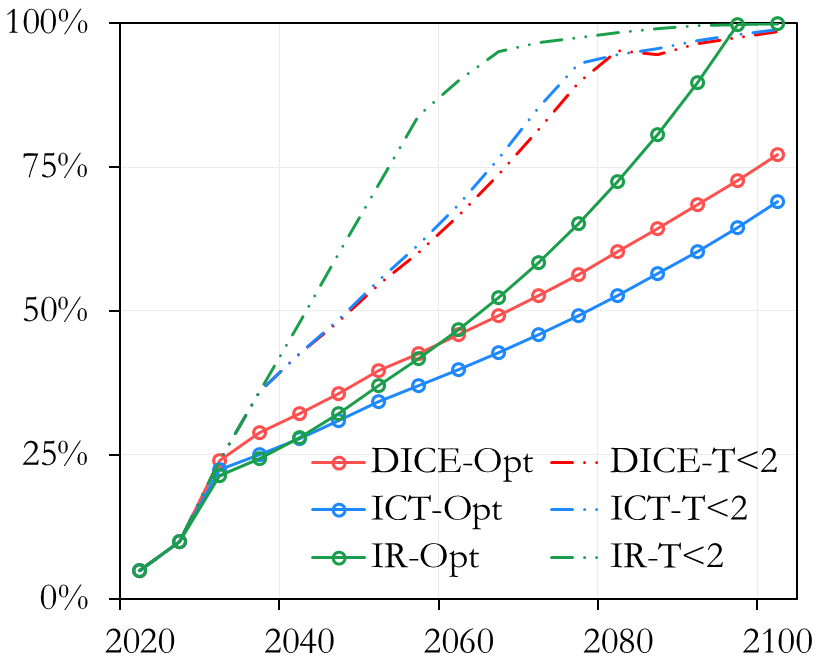}
        \caption{With adaptation}
        \label{ef2b}
    \end{subfigure}
    \hfill
    \begin{subfigure}[b]{0.4\textwidth}
        \centering
        \includegraphics[width=\textwidth]{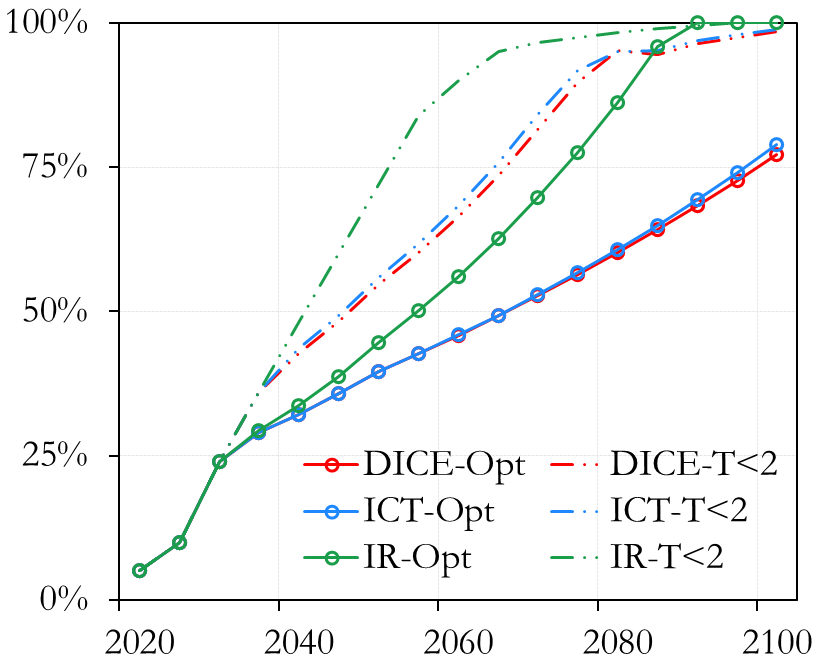}
        \caption{Without adaptation}
        \label{ef2b}
    \end{subfigure}
    
    \caption{Optimal abatement rate under different economic prospects}
    \label{ef2}

    \vspace{2mm}
    \begin{minipage}{\textwidth}
        \footnotesize
        \textbf{Note:} ICT projects a modest influence on long-run growth from AI, much like earlier information and communications technologies. IR posits a transformative economic impact sufficient to trigger a new industrial revolution. OPT employs an optimal carbon price to maximize social welfare. T$<$2 caps the global mean temperature rise at 2\degree C above the preindustrial level.
    \end{minipage}
    
\end{figure}

\subsection{AI-related welfare analysis}

\begin{sidewaystable}[htbp]
  \centering
  \caption{Welfare analysis of AI under different climate policies}
    \begin{tabular}{cclcccccc}
    \toprule
    \multicolumn{3}{c}{\multirow{2}[4]{*}{Scenario}} & \multicolumn{2}{c}{AI-related benefits} & \multicolumn{2}{c}{With AI-powered adaptation} & \multicolumn{2}{c}{Without  AI-powered adaptation} \\
\cmidrule{4-9}    \multicolumn{3}{c}{}  & \multicolumn{1}{p{7.5em}}{Economic gains} & \multicolumn{1}{p{7.5em}}{\quad Full gains} & \multicolumn{1}{p{7.5em}}{Climate loss} & \multicolumn{1}{p{7.5em}}{\qquad Offset} & \multicolumn{1}{p{7.5em}}{Climate loss} & \multicolumn{1}{p{7.5em}}{\qquad Offset} \\
    \midrule
    \multirow{12}[4]{*}{BAU} & \multirow{6}[2]{*}{ICT} & Baseline & 7.07\% & 7.62\% & 1.59\% & 20.82\% & 2.73\% & 35.77\% \\
          &       & Philosopher & 8.06\% & 8.61\% & 1.86\% & 21.58\% & 3.41\% & 39.55\% \\
          &       & Stern & 46.88\% & 47.27\% & 14.96\% & 31.65\% & 19.46\% & 41.16\% \\
          &       & Nordhaus & 3.67\% & 4.16\% & 0.70\% & 16.73\% & 1.59\% & 38.21\% \\
          &       & Alt damage 1 & 6.98\% & 7.57\% & 0.40\% & 5.29\% & 1.27\% & 16.83\% \\
          &       & Alt damage 2 & 6.87\% & 14.74\% & 15.60\% & 105.86\% & 26.35\% & 178.79\% \\
\cmidrule{2-9}          & \multirow{6}[2]{*}{IR} & Baseline & 85.92\% & 87.18\% & 23.24\% & 26.66\% & 25.73\% & 29.51\% \\
          &       & Philosopher & 92.68\% & 93.99\% & 25.22\% & 26.84\% & 27.88\% & 29.66\% \\
          &       & Stern & 918.49\% & 922.38\% & 417.79\% & 45.29\% & 458.57\% & 49.72\% \\
          &       & Nordhaus & 48.97\% & 49.84\% & 8.98\% & 18.02\% & 10.47\% & 21.00\% \\
          &       & Alt damage 1 & 86.17\% & 87.44\% & 21.92\% & 25.06\% & 23.31\% & 26.66\% \\
          &       & Alt damage 2 & 89.33\% & 104.92\% & 47.69\% & 45.45\% & 74.10\% & 70.62\% \\
    \midrule
    \multirow{12}[4]{*}{Opt} & \multirow{6}[2]{*}{ICT} & Baseline & 7.07\% & 7.62\% & 0.28\% & 3.64\% & 0.58\% & 7.57\% \\
          &       & Philosopher & 8.06\% & 8.61\% & 0.30\% & 3.45\% & 0.89\% & 10.29\% \\
          &       & Stern & 46.88\% & 47.27\% & 0.34\% & 0.71\% & 0.78\% & 1.64\% \\
          &       & Nordhaus & 3.67\% & 4.16\% & 0.20\% & 4.71\% & 0.70\% & 16.94\% \\
          &       & Alt damage 1 & 6.98\% & 7.57\% & 0.19\% & 2.52\% & 0.79\% & 10.48\% \\
          &       & Alt damage 2 & 6.87\% & 14.74\% & 1.17\% & 7.97\% & 8.82\% & 59.81\% \\
\cmidrule{2-9}          & \multirow{6}[2]{*}{IR} & Baseline & 85.92\% & 87.18\% & 0.99\% & 1.14\% & 2.47\% & 2.83\% \\
          &       & Philosopher & 92.68\% & 93.99\% & 1.10\% & 1.17\% & 2.63\% & 2.80\% \\
          &       & Stern & 918.49\% & 922.38\% & 3.03\% & 0.33\% & 7.45\% & 0.81\% \\
          &       & Nordhaus & 48.97\% & 49.84\% & 0.67\% & 1.35\% & 1.73\% & 3.47\% \\
          &       & Alt damage 1 & 86.17\% & 87.44\% & 0.83\% & 0.95\% & 2.33\% & 2.66\% \\
          &       & Alt damage 2 & 89.33\% & 104.92\% & 4.12\% & 3.93\% & 20.31\% & 19.36\% \\
    \midrule
    \multirow{12}[4]{*}{T$<2$\degree C } & \multirow{6}[2]{*}{ICT} & Baseline & 7.07\% & 7.62\% & 0.81\% & 10.60\% & 0.91\% & 11.97\% \\
          &       & Philosopher & 8.06\% & 8.61\% & 0.81\% & 9.35\% & 1.20\% & 13.91\% \\
          &       & Stern & 46.88\% & 47.27\% & 0.53\% & 1.12\% & 0.89\% & 1.89\% \\
          &       & Nordhaus & 3.67\% & 4.16\% & 0.81\% & 19.44\% & 1.14\% & 27.36\% \\
          &       & Alt damage 1 & 6.98\% &7.57\% & 1.19\% & 15.68\% & 1.55\% & 20.53\% \\
          &       & Alt damage 2 & 6.87\% & 14.74\% & 1.17\% & 7.97\% & 8.82\% & 59.81\% \\
\cmidrule{2-9}          & \multirow{6}[2]{*}{IR} & Baseline & 85.92\% & 87.18\% & 2.60\% & 2.98\% & 3.48\% & 3.99\% \\
          &       & Philosopher & 92.68\% & 93.99\% & 2.78\% & 2.96\% & 3.70\% & 3.94\% \\
          &       & Stern & 918.49\% & 922.38\% & 32.40\% & 3.51\% & 35.20\% & 3.82\% \\
          &       & Nordhaus & 48.97\% & 49.84\% & 1.93\% & 3.87\% & 2.59\% & 5.19\% \\
          &       & Alt damage 1 & 86.17\% & 87.44\% & 3.61\% & 4.13\% & 4.55\% & 5.09\% \\
          &       & Alt damage 2 & 89.33\% & 104.92\% & 4.12\% & 3.93\% & 20.31\% & 19.36\% \\
    \bottomrule
    \end{tabular}%
  \label{et01}%

\begin{tablenotes}
        \footnotesize
        \item \textbf{Note:} Using permanent-equivalent-variation method following Ref. \cite{barrage2020optimal}, the figure displays welfare changes from AI development, expressed relative to the consumption path of the DICE-2023 optimal scenario. 'Full gain' consists of both the AI 'Economic gains' and AI-powered climate adaptation. The climate-related loss captures both climate damages from AI-induced warming and associated abatement costs (see 'Methods'). 'Offset' is calculated as a share of 'Full gains'. ICT projects a modest influence on long-run growth from AI, much like earlier information and communications technologies. IR posits a transformative economic impact sufficient to trigger a new industrial revolution. BAU represents the current business-as-usual low abatement policy as implemented in DICE-2023. OPT employs an optimal carbon price to maximize social welfare. $T<2$\degree C caps the global mean temperature rise at 2\degree C above the preindustrial level. The 'No adaptation' scenario precludes AI-powered adaptation from reducing climate damages, as opposed to the Baseline scenario, which assumes that AI development fulfills half of its adaptation potentials by 2050. Our baseline simulation adopts the discount rate parameters based on a survey of economists \cite{drupp2018discounting}. For sensitivity tests, we also consider discounting parameters advocated by philosophers \cite{nesje2023philosophers}, used in the Stern Report \cite{stern2007economics} and in earlier work by Nordhaus \cite{nordhaus2018projections}. Additionally, we include two alternative damage estimates from a recent updated meta-analysis of climate damages \cite{Tol2024ClimateMeta}.
    \end{tablenotes}

\end{sidewaystable}%

\newpage

\subsection{Can AI be non net polluting?}

\begin{figure}[htbp]
    \centering
    \includegraphics[width=0.66\linewidth]{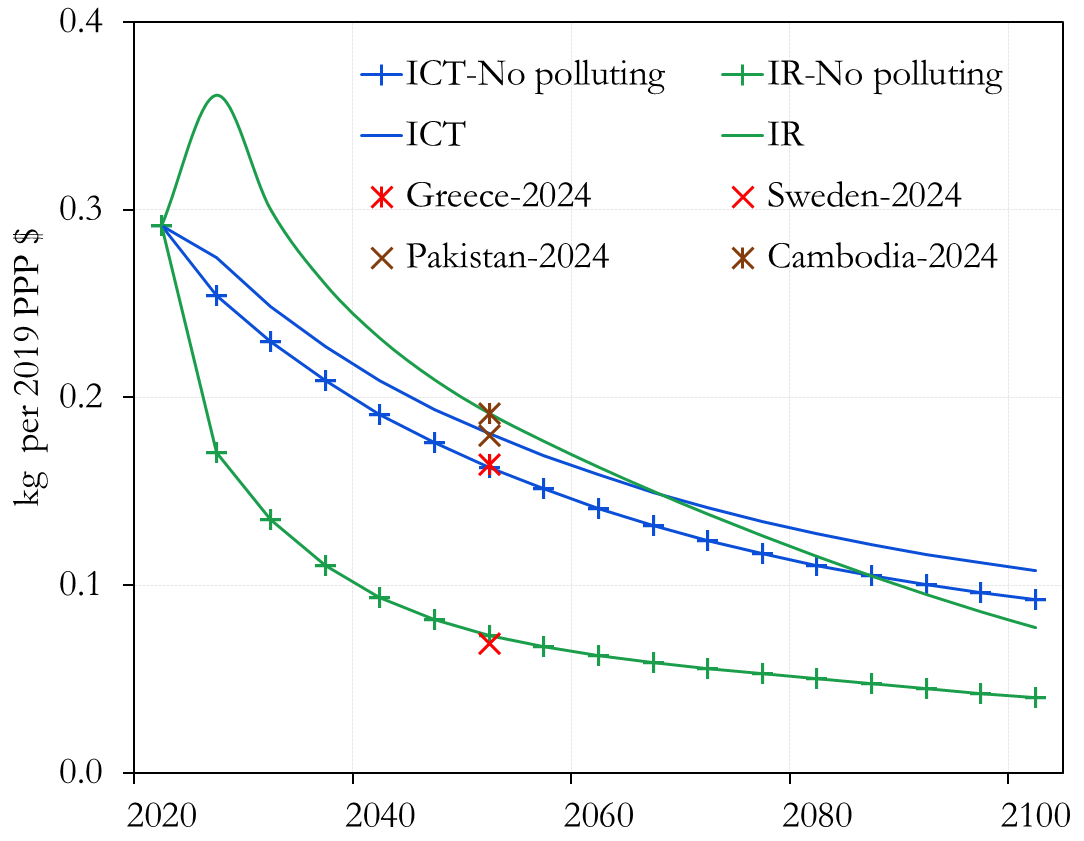}
    \caption{Production-based carbon intensity under the thought experiment}
    \label{ef03}
    
    \vspace{2mm}
    \begin{minipage}{\textwidth}
        \footnotesize
        \textbf{Note:}  ICT projects a modest influence on long-run growth from AI, much like earlier information and communications technologies. IR posits a transformative economic impact sufficient to trigger a new industrial revolution. Emissions from AI and general capital are aggregated and then divided by aggregate final output, yielding the standard production-based carbon intensity as used in DICE-2023. In the figure, solid curves depict the time path of this carbon intensity under the BAU low abatement scenario in the DICE-AI model. The solid line with plus markers shows the counterfactual time path required for AI’s abatement potential to fully offset emissions from its computing-related and economy-wide impacts. Country-specific carbon intensity data are sourced from the World Bank World Development Indicators and harmonized with the initial global carbon intensity in DICE-2023.
    \end{minipage}
    
\end{figure}

\newpage

\subsection{Sensitivity analysis}\label{sensi_ex}

\begin{figure}[htbp]
    \centering
    \includegraphics[width=1 \linewidth]{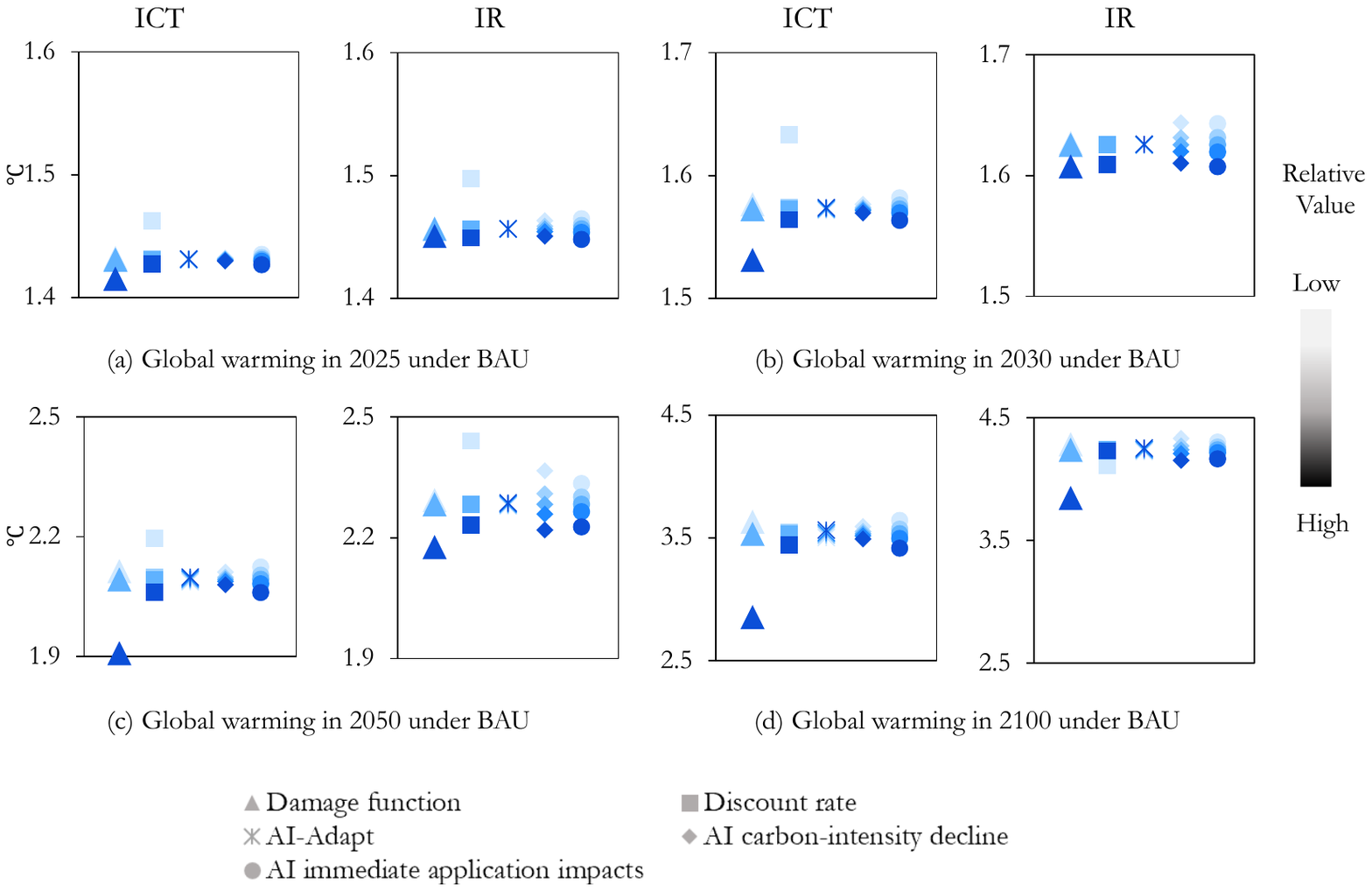}
    \caption{Global warming }
    \label{ex_5a}
 \vspace{2mm}
    \begin{minipage}{\textwidth}
        \footnotesize
  \textbf{Note:} ICT projects a modest influence on long-run growth from AI, much like earlier information and communications technologies. IR posits a transformative economic impact sufficient to trigger a new industrial revolution. The darker the mark, the higher the value it represents. For the damage function, we consider three values in the sensitivity analysis: our baseline specification and two alternatives from \cite{Tol2024ClimateMeta}. For the discount rate, we consider four values: our baseline value (the mean recommended by economists, \cite{drupp2018discounting}), the mean recommended by philosophers, \cite{nesje2023philosophers}, and the values used by \cite{stern2007economics} and \cite{nordhaus2018projections}. For AI-powered adaptation ($a_{3}$), the AI carbon-intensity decline rate (captured by the ratio of the AI-sector decline rate to that of the final production sector in DICE-2023), and AI immediate application impacts ($\psi_{adj}$), we first determine the lower and upper bounds of each parameter based on literature review, then assume a normal distribution with the baseline value as the mean, and finally select the values at the 2.5th, 25th, 50th, 75th, and 97.5th percentiles for the sensitivity analysis. Further details on these parameters are provided in the Calibration section.
    \end{minipage}
    
\end{figure}
\begin{figure}[htbp]
    \centering
    \includegraphics[width=1 \linewidth]{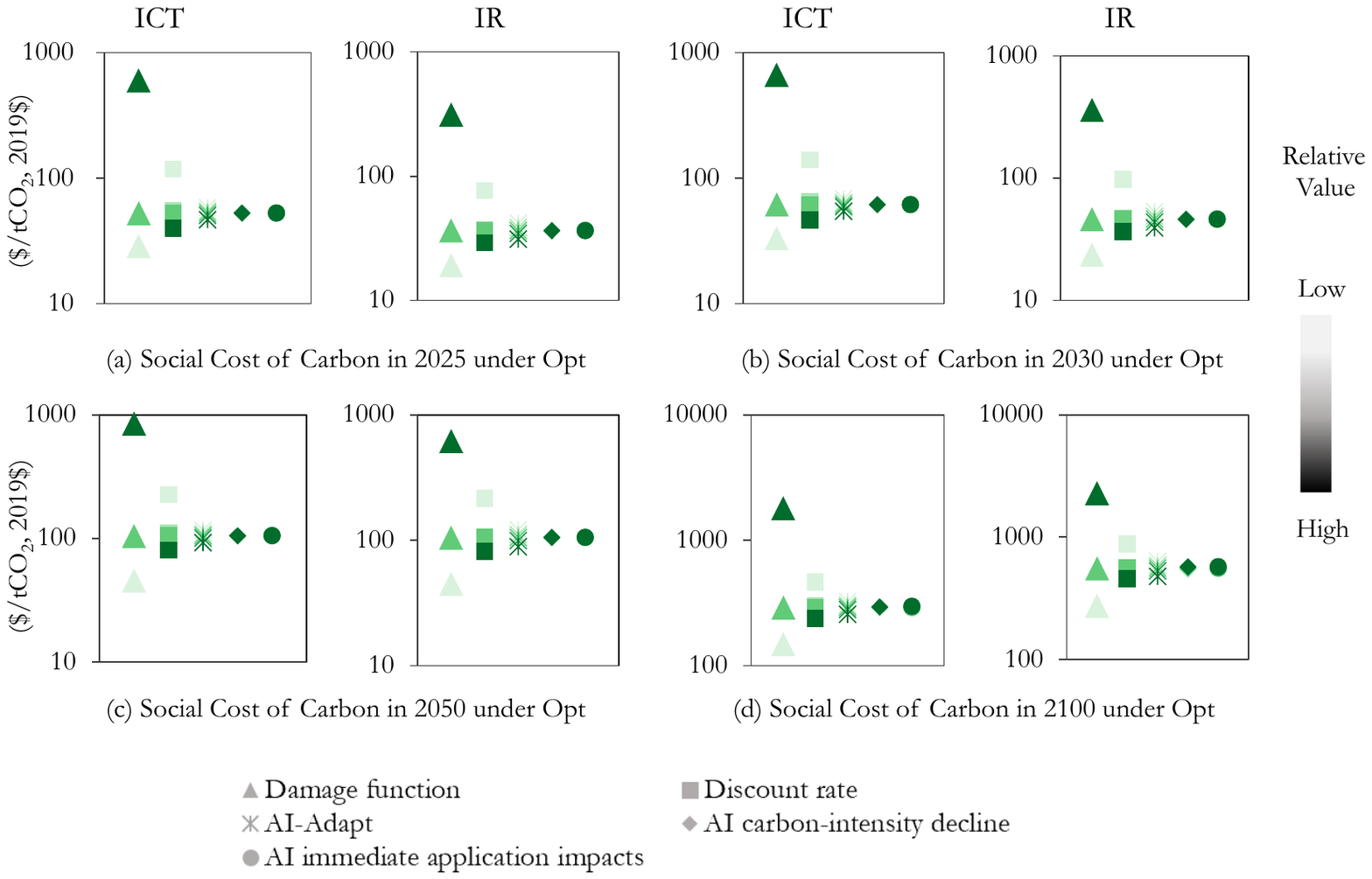}
    \caption{Social Cost of Carbon }
    \label{ex_5b}
 \vspace{2mm}
    \begin{minipage}{\textwidth}
        \footnotesize
  \textbf{Note:} ICT projects a modest influence on long-run growth from AI, much like earlier information and communications technologies. IR posits a transformative economic impact sufficient to trigger a new industrial revolution. The darker the mark, the higher the value it represents. For the damage function, we consider three values in the sensitivity analysis: our baseline specification and two alternatives from \cite{Tol2024ClimateMeta}. For the discount rate, we consider four values: our baseline value (the mean recommended by economists, \cite{drupp2018discounting}), the mean recommended by philosophers, \cite{nesje2023philosophers}, and the values used by \cite{stern2007economics} and \cite{nordhaus2018projections}. For AI-powered adaptation ($a_{3}$), the AI carbon-intensity decline rate (captured by the ratio of the AI-sector decline rate to that of the final production sector in DICE-2023), and AI immediate application impacts ($\psi_{adj}$), we first determine the lower and upper bounds of each parameter based on literature review, then assume a normal distribution with the baseline value as the mean, and finally select the values at the 2.5th, 25th, 50th, 75th, and 97.5th percentiles for the sensitivity analysis. Further details on these parameters are provided in the Calibration section.
    \end{minipage}
    
\end{figure}

\begin{figure}[htbp]
    \centering
    \includegraphics[width=1 \linewidth]{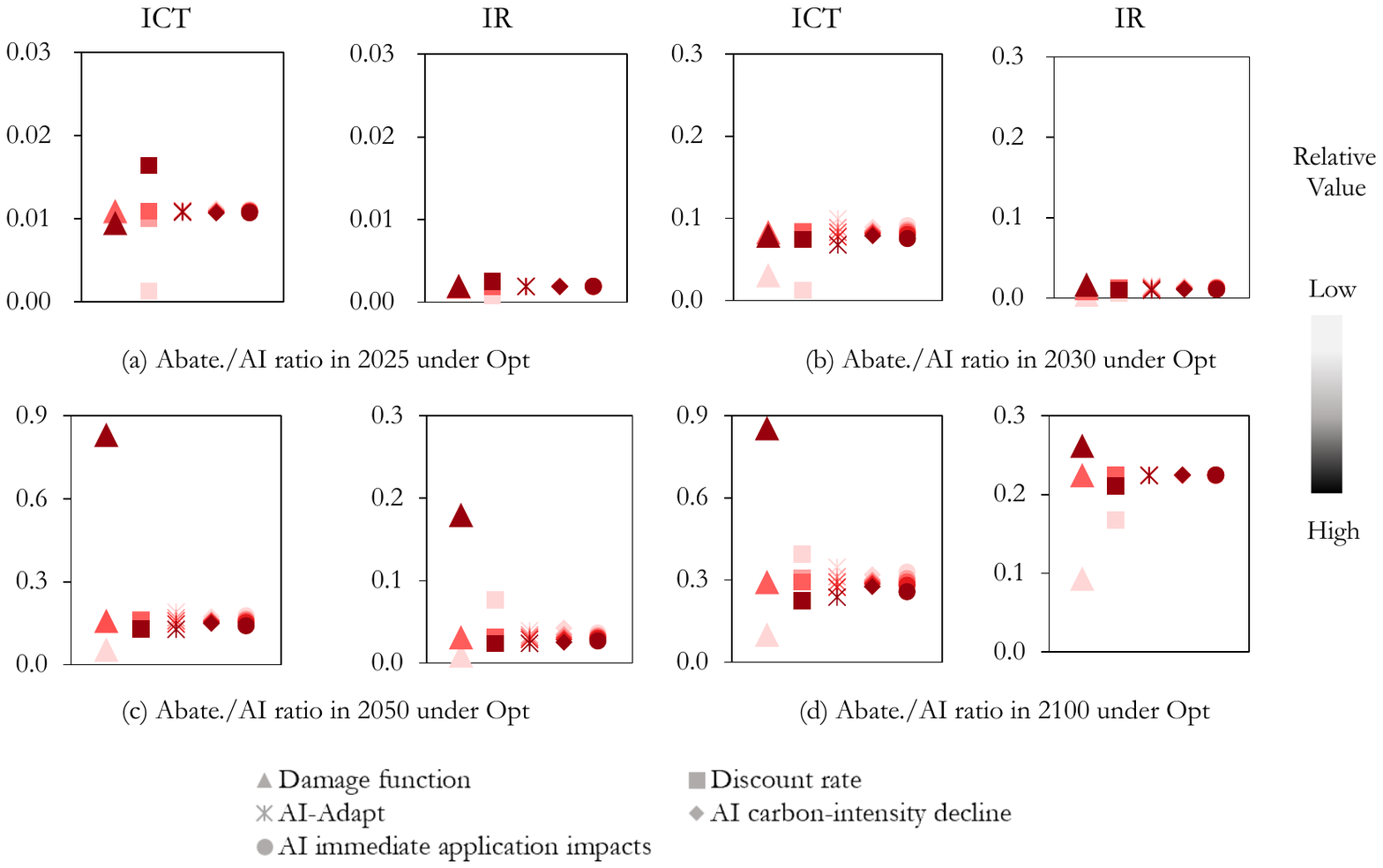}
    \caption{Abate./AI ratio}
    \label{ex_5c}
 \vspace{2mm}
    \begin{minipage}{\textwidth}
        \footnotesize
  \textbf{Note:} ICT projects a modest influence on long-run growth from AI, much like earlier information and communications technologies. IR posits a transformative economic impact sufficient to trigger a new industrial revolution. The darker the mark, the higher the value it represents. For the damage function, we consider three values in the sensitivity analysis: our baseline specification and two alternatives from \cite{Tol2024ClimateMeta}. For the discount rate, we consider four values: our baseline value (the mean recommended by economists, \cite{drupp2018discounting}), the mean recommended by philosophers, \cite{nesje2023philosophers}, and the values used by \cite{stern2007economics} and \cite{nordhaus2018projections}. For AI-powered adaptation ($a_{3}$), the AI carbon-intensity decline rate (captured by the ratio of the AI-sector decline rate to that of the final production sector in DICE-2023), and AI immediate application impacts ($\psi_{adj}$), we first determine the lower and upper bounds of each parameter based on literature review, then assume a normal distribution with the baseline value as the mean, and finally select the values at the 2.5th, 25th, 50th, 75th, and 97.5th percentiles for the sensitivity analysis. Further details on these parameters are provided in the Calibration section.
    \end{minipage}
    
\end{figure}
\clearpage

\section{Literature review}\label{secA}

We collect the literature on three perspectives of AI's impact, and examines their implications for climate change and climate policy, specifically through the lenses of economy, energy and climate. This categorization slightly differs from the framework presented in Fig.\ref{fig01}, which instead separates economy, emission, and climate. The adjustment is motivated by the observation that energy represents a prominent and widely discussed aspect of AI’s influence in both the literature and policy discussions \cite{masanet2020recalibrating,kaack2022aligning,IEA2025EnergyAI,shehabi20242024}. Several clarifications can be useful. The 'emission' impact encompasses the 'energy' impact, because AI influence emissions through other channels, such as lifecycle emissions from building AI-related infrastructure and its immediate application impacts. In addition, AI’s economic and emission impacts are partially overlapping, since AI-driven economic expansion indirectly raises emissions by enlarging the production base. 

Given the breadth of topics intersecting AI and climate, a comprehensive review of all relevant literature is beyond the scope of this study and is hardly possible. Our primary aim is to explore the key interactions between AI and climate change, thereby positioning the present study and clarifying its contributions. For reviews on AI' economic impact, we refer to Ref.\cite{14015} and Ref.\cite{trammell2023economic}; for reviews on AI' emission impact, we refer to Ref.\cite{kaack2022aligning}; for reviews on AI's use in climate change, we refer to Ref.\cite{rolnick2022tackling}. The core methodology in this paper builds on integrated assessment modelling, including both aggregate and detailed process IAMs, as reviewed in Refs. \cite{weyant2017some,bosetti2021integrated,dietz2024introduction}.

First, we review the economic impact of AI (Table \ref{tab_ap01}), which is relevant for both economic performance and emissions. This impact operates through two primary channels, productivity growth and capital deepening. By enhancing total factor productivity, AI raises aggregate economic output. This expansion not only permits resources for consumption and investment but also creates a larger base for carbon emissions. At the same time, AI drives capital deepening, whereby capital progressively substitutes for labor in production. While this boosts economic output, it also potentially introduces adverse side effects, including higher emissions and greater inequality.

\begin{sidewaystable}[htbp]
  \centering
  \caption{The economic impact of AI and its implications for climate change and climate policy}
    \begin{tabular}{clp{23.9em}p{28.6em}}
    \toprule
          & \multicolumn{1}{p{4.585em}}{Aspect} & Details & Discussions \\
    \midrule
    \multicolumn{1}{c}{\multirow{5}[10]{*}{Economy}} & \multicolumn{1}{l}{\multirow{3}[6]{*}{TFP gains}} & AI acts as both a labor-augmenting technology (for instance, assisting an inexperienced driver through an automated system), and as a catalyst for innovation, generating new products, patents, and knowledge \cite{14015,trammell2023economic,erdil2023explosive,acemoglu2025simple}. For example, it accelerates R\&D (e.g., Nobel Prize in Chemistry for AI use in protein folding \cite{aaqvist2024computational}) and fuels firm growth through product innovation \cite{babina2024artificial}. Notably, both technology workers and the firms that employ them are more exposed to AI \cite{eloundou2024gpts,labaschin2025extending}.  & Despite broad agreement that AI enhances productivity, projections of its economic magnitude vary dramatically across academia \cite{davidson2021could,acemoglu2025simple,aghion2024ai} and major institutions \cite{chui2023state,filippucci2024impact,cerutti2025global}. Whether its impact will be moderate or transformative enough to launch a new industrial revolution remains uncertain (e.g., the Baumol effect \cite{jones2024framework,jonespast}). Although several AI scenarios have been examined in economic research \cite{korinek2023scenario,trammell2023economic}, they have not yet been integrated into different socioeconomic pathways for climate-related studies \cite{luers2024will}. \\
\cmidrule{3-4}          &       & The scale of production fundamentally shapes greenhouse gas emissions \cite{rennert2022comprehensive}. Indeed, economic projections serve as key inputs to integrated assessment models (IAMs) \cite{nordhaus2018projections} and Shared Socioeconomic Pathways (SSPs) \cite{dellink2017long,o2017roads}. A larger economic base typically leads to higher carbon emissions.    & Economic prosperity could also support emission reductions, provided that advances in general-purpose technologies spur innovation in abatement and channel resources toward low-carbon activities \cite{hepburn2013prosperity,iea2025}. Given AI’s potential to drive future affluence, its net emission impact remains ambiguous, hinging on both the evolving size of the economy and resource allocation alongside climate policy.  \\
\cmidrule{3-4}          &      &  On the welfare, an expanded production base supports greater consumption and investment. Higher consumption improves current well-being,  while increased investment in productive capital boosts future consumption potential, enhancing future welfare \cite{barro2025economic,acemoglu2008introduction}. & If a larger AI-augmented economy raises emissions, it accordingly elevates environmental costs \cite{Korinek2025EconomicAI}, which weigh against above benefits. Yet such trade-offs have rarely been quantified in the literature. As a result, the net welfare impact of AI remains uncertain, which should be assessed with other aspects of AI \cite{Korinek2025EconomicAI,jones2024ai}. \\
\cmidrule{2-4}          & \multicolumn{1}{l}{\multirow{2}[4]{*}{Capital deepening}} & From an emissions perspective, energy consumption is closely linked to capital formation \cite{struckmeyer1987putty,atkeson1999models,baldwin2020build,van2020stranded,campiglio2022optimal}. Capital deepening, driven by increased capital-labor substitutability \cite{nordhaus2021we} or task automation \cite{zeira1998workers,acemoglu2018race,jones2024framework} due to AI, implies a greater reliance on machinery to replace human labor \cite{trammell2023economic,acemoglu2025simple}. When additional equipment, machinery, or factories are deployed, they typically require more energy to operate, thereby raising emissions.  & While several recent studies have identified channels through which AI may raise or lower carbon emissions \cite{kaack2022aligning,IEA2025EnergyAI,stern2025green,lu2025influence}, less attention has been paid to the direct emissions increase from operating more capital (the \emph{intensive} margin of capital deepening). This can bias the assessment of AI's net environmental impact. A recent study \cite{harding2025watts} evaluates the direct emission impacts of AI adoption, yet its analyses does not incorporate broader general equilibrium effects. \\
\cmidrule{3-4}          &       & The welfare effects of AI-driven capital deepening are multifaceted. On one hand, a rising capital share in production enhances economic output \cite{trammell2023economic,acemoglu2025simple,filippucci2024impact} and can thereby support higher consumption. On the other hand, as AI substitutes capital for labor, the labor income share tends to decline in national accounts \cite{acemoglu2018race,acemoglu2022tasks,grossman2022elusive}. A falling labor share, accompanied by rising capital share, is often linked to widening inequality \cite{piketty2014capital}, which may adversely affect overall social welfare.  &  Capital deepening also affects emissions on the \emph{extensive} margin, when it boosts all polluting economic activities at a larger economic base, other than emissions directly from using more capital (\emph{intensive} margin). These dynamic impacts of AI are not yet captured in current IAMs or climate policy studies \cite{stern2025green,luers2024will}. Although AI also raises distributional concerns that affect welfare \cite{korinek2021artificial,10.1093/oxfordhb/9780197579329.013.44,ren2024uneven}, such inequality effects are difficult to analyze within highly aggregated models like DICE \cite{emmerling2024multi}. Examining AI’s impact on inequality thus remains an important direction for future research.  \\
    \bottomrule
    \end{tabular}%
  \label{tab_ap01}%
\end{sidewaystable}%

Table \ref{tab_ap02} summarizes the literature on AI's energy and climate implications. Existing studies considers both the energy-intensive nature of AI development and its potential to improve energy efficiency.  In addition, AI also influences emissions through channels not directly linked to energy use. Consequently, the net emissions impact of AI remains ambiguous and must be evaluated by integrating all these aspects together under different mitigation policy scenarios within a consistent analytical framework. Ultimately, the combined emission and adaptation effects of AI will shape its environmental costs and determine the optimal level of carbon taxation. A formal assessment, however, requires an integrated climate–economy–AI modeling framework.

\begin{sidewaystable}[htbp]
  \centering
  \caption{The energy and climate impacts of AI and their implications for climate change and climate policy}
    \begin{tabular}{cp{6em}p{22em}p{30.5em}}
    \toprule
          & Aspect & Details & Discussions \\
    \midrule
    \multicolumn{1}{c}{\multirow{2}[4]{*}{Energy}} & Energy-intensive AI &  The training, inference, and development of AI systems are computationally intensive, significantly increasing energy demand \cite{de2023growing,kaack2022aligning,masanet2020recalibrating,bogmans2025power}. This demand is further amplified when considering the full lifecycle of AI infrastructure, including the manufacturing, transportation, and disposal of related hardware \cite{ITU2025AIEnvironment,itu_wba_greening_2025}. Collectively, these constitute the computing-related energy impacts of AI. &  While recent studies have attempted to quantify AI-related energy use, these estimates are subject to substantial uncertainty, due to limited transparency and data availability of AI data center \cite{masanet2024better}, inconsistent measurement methodologies \cite{ITU2025AIEnvironment}, geographic variability \cite{gibney2022shrink}, etc. Moreover, the future carbon intensity of AI remains highly uncertain, shaped by countervailing forces. On one hand, factors such as wider AI adoption, rebound effects that raise energy demand, and the increasing scale of models requiring more training data may elevate emissions \cite{masanet2020recalibrating,Bashir2024Climate}. On the other, improvements in hardware and algorithmic efficiency, together with a growing share of renewable energy in powering AI operations, could reduce the carbon footprint \cite{masanet2020recalibrating,de2023growing,IEA2025EnergyAI}. \\
\cmidrule{2-4}          & Energy-saving AI & By redesigning and transforming complex systems \cite{stern2025green}, AI can optimize energy supply, increase the transportation efficiency in electricity networks, improve energy end-use, etc \cite{Gentine2024AIClimateNature,IEA2025EnergyAI}. Also, AI can be expected to accelerate battery innovation for energy storage, catalysts in synthetic fuel, etc \cite{IEA2025EnergyAI}. & Some recent studies are positive and speculate that the potential of AI to abate emissions will outweigh the additional emissions of energy consumption in data centers \cite{stern2025green}. However, these estimates are subject to significant uncertainties and have not yet included the economy-wide emissions, which are critical for a complete assessment of AI’s emission impact \cite{kaack2022aligning,luers2024will}. \\
                                                   \midrule
    \multicolumn{1}{c}{\multirow{2}[4]{*}{Climate}} & Non-energy mitigation & AI also influences emissions that are not directly tied to energy use. For example, it can aid in identifying carbon sinks, innovating carbon capture materials, and updating clinker reduction methods in cement production \cite{gloege2022improved,asibor2023machine,IEA2025EnergyAI}. Furthermore, AI can support low‑carbon shifts in consumer behavior, such as through smart home systems with learning capabilities or AI‑guided planning of more fuel‑efficient travel routes \cite{stern2025green}.  & Together with its potential to reduce direct energy demand, these channels represent positive environmental impacts of AI and are expected to support progress toward climate targets. Both preference‑driven behavior changes \cite{stern2022economics,besley2023political} and negative‑emission technologies play important roles in deep‑decarbonization scenarios. However, current estimates of AI’s emission‑reduction potential vary widely \cite{luderer2018residual,wei2021proposed}. Besides, future research could develop IAMs that account for how AI shapes social norms and preferences, thus informing both the pace and policy design in the green transition \cite{stern2025green,wang2024endogenous}. \\
\cmidrule{2-4}          & Climate adaptation  & AI can empower next-generation multiscale climate modeling \cite{eyring2024ai}, improving the monitoring, forecasting, and warning systems for climate-related events \cite{Gentine2024AIClimateNature}. For instance, by integrating physical laws with machine learning, models such as GraphCast and Google's FloodHub enhance the accuracy of weather and flood predictions \cite{lam2023learning,nearing2024global}. It may also foster developments in solar geoengineering \cite{rolnick2022tackling}.  & Even without considering AI's impact \cite{CARLETON2024143}, quantifying the potential of climate adaptation is inherently difficult. Despite the promising future of AI adoption in climate adaptation \cite{rolnick2022tackling,BibriCoi2025}, the extent to which it can reduce climate impacts remains uncertain and is not yet backed by empirical evidence. When AI-powered adaptation takes effect, its interaction with AI's emission impacts alters the associated climate costs, thereby shifting the optimal climate policy derived from cost-benefit analysis.\\
    \bottomrule
    \end{tabular}%
  \label{tab_ap02}%
\end{sidewaystable}%

\newpage

\section{Model}\label{secB}

\begin{figure}[htbp]
    \centering
    \includegraphics[width=1\linewidth]{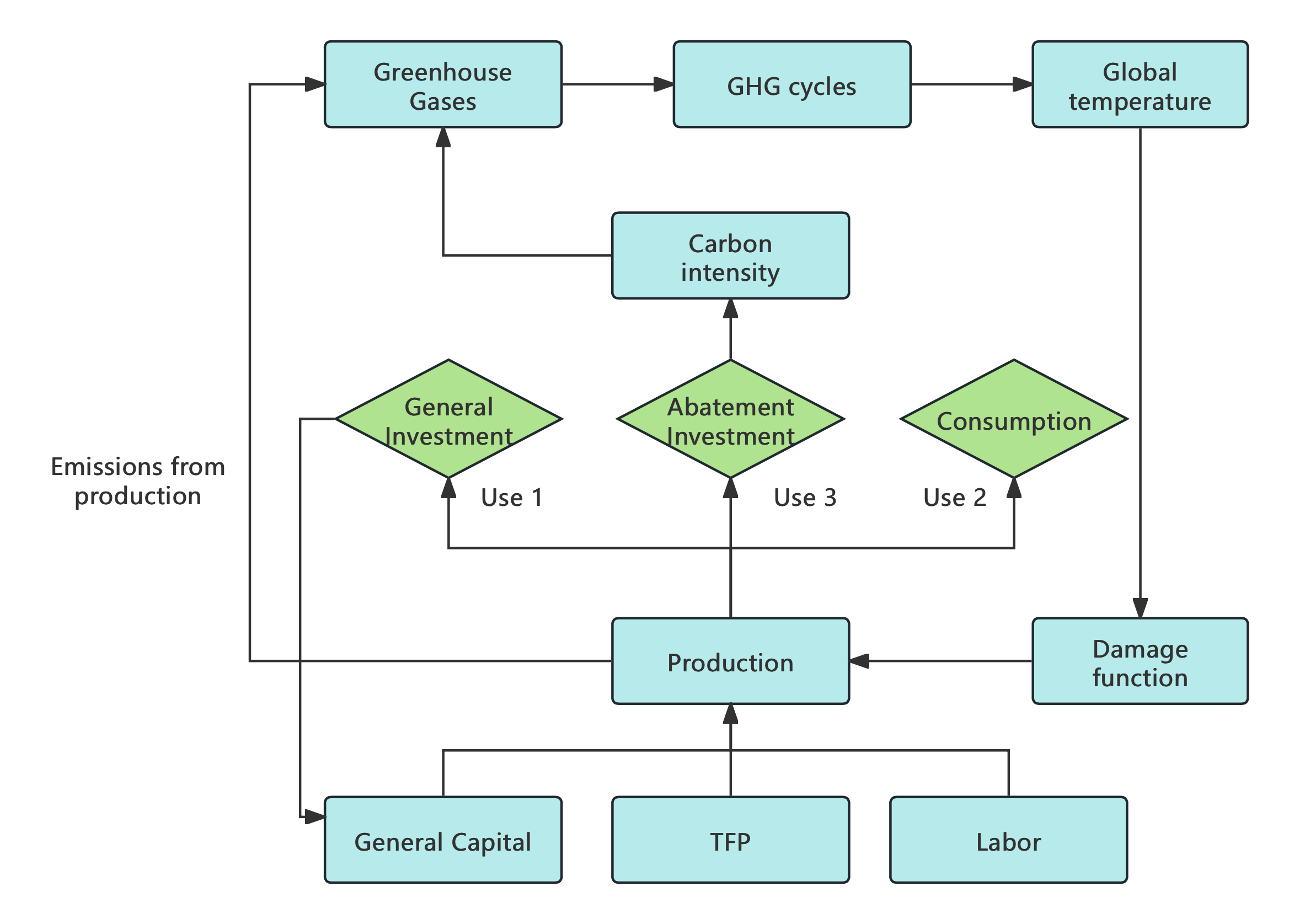}
    \caption{The framework of DICE-2023}
    \label{fig_ap01}
\end{figure}

The framework of the DICE-2023 model is illustrated in Fig. \ref{fig_ap01}, with further technical details available in Ref. \cite{barrage2024policies}. a global social planner determines consumption, general investment, and abatement investment in each 5‑year period, starting in 2020. The planner maximizes social welfare, which depends solely on the time path of consumption. Although recent studies have extended the utility function to include non‑consumption arguments \cite{barrage2020optimal,drupp2021relative,bastien2021use,wang2024endogenous}, we adhere to the DICE‑2023 specification to maintain focus within the analytical framework outlined in Fig. \ref{fig01}.

\begin{figure}[htbp]
    \centering
    \includegraphics[width=1.2\linewidth]{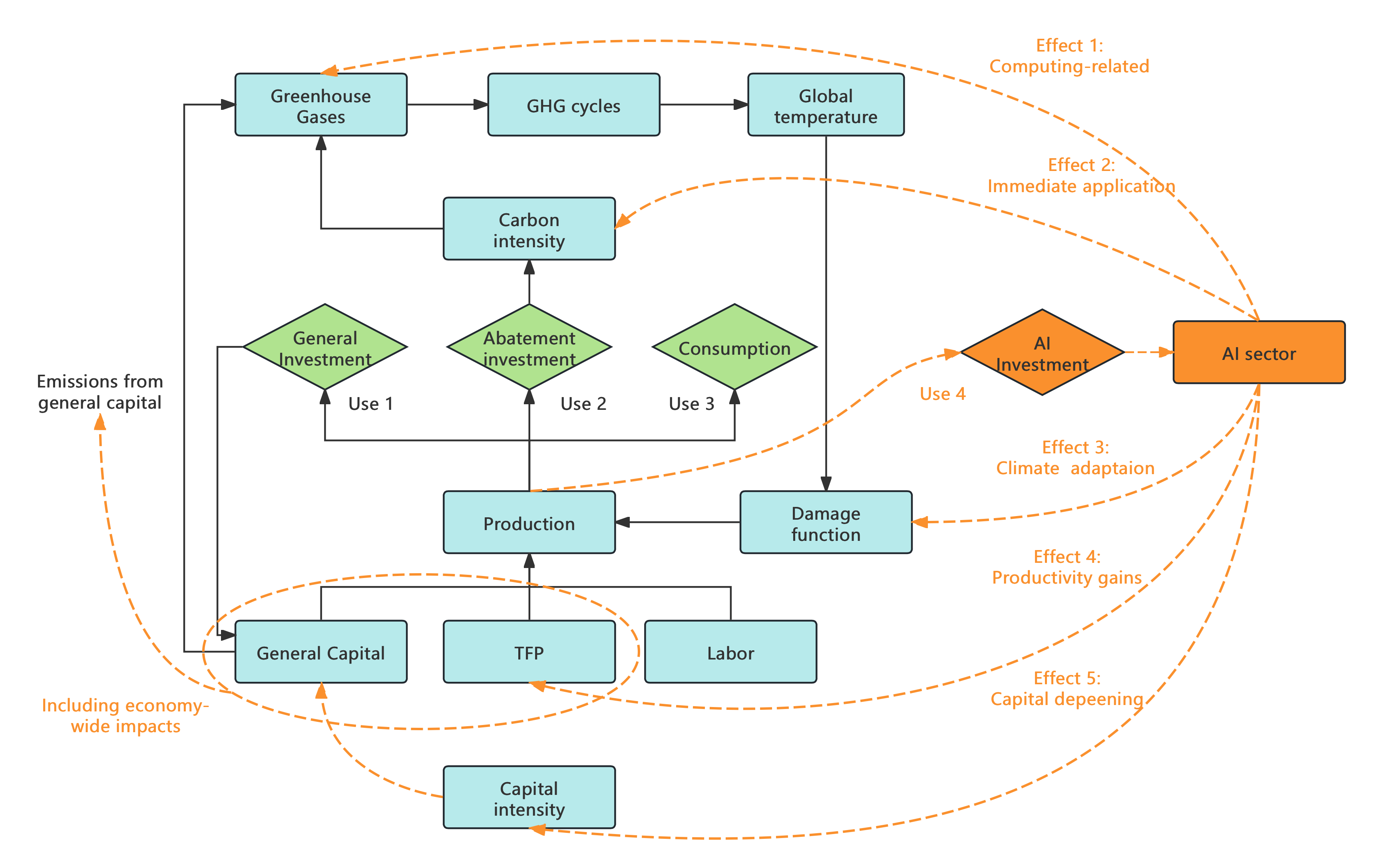}
    \caption{The framework of DICE-AI}
    \label{fig_ap02}
\end{figure}

Our extended model, DICE‑AI, modifies this structure in two key respects (Fig. \ref{fig_ap02}). First, it shifts from a production‑based to a capital‑embodied accounting of carbon intensity for emissions calculation (see Appendix \ref{ap_cal_em}). Second, it integrates the three perspectives of AI outlined in our conceptual framework (Fig. \ref{fig01}) directly into the DICE model structure. 

\newpage

\section{Calibration}\label{secC}

Our calibration strategy seeks to maintain the best comparability with the DICE-2023 model, thereby isolating how the introduction of AI alters its climate‑economic projections and associated policy implications.  For parameters common to both models, we retain the original DICE-2023 values, making adjustments only where necessary to ensure consistency after incorporating an intermediate AI production sector. Parameters related to the AI sector are calibrated through an extensive review of the existing literature, including the AI production function, AI-induced effects on economic growth, the carbon intensity of AI-related capital, AI-induced mitigation, and AI-powered climate adaptation.

\subsection{Utility function and discount rate}
The DICE-AI model accommodates a representative household whose instantaneous utility, $U_t$, depends exclusively on per capita consumption, $c_t$. The utility function adopts the standard constant‑relative‑risk‑aversion (CRRA) form with elasticity $\eta$. DICE-AI maximizes the net present value of aggregate population-weighted utility over time:
\begin{align}
\label{eq:utility}
\max V &= \sum_{t=1}^{T} U_t(c_t) L_t^{total} R_t\notag\\
    &= \sum_{t=1}^{T} \frac{c_t^{\,1-\eta}}{1-\eta} L_t^{total}  (1+\rho)^{-t}
\end{align}
where $L_t^{total}$ denotes labor input whose projected values are taken from DICE-2023 model. $R_t$ is the time-specific consumption discount factor and $\rho$ the constant social time preference rate.

DICE-2023 adopts a time-varying discount rate, to reflect both the precautionary effect \cite{gollier2014discounting,newell2022discounting} and the climate beta effects \cite{dietz2018climate}. While these refinements are conceptually justified, they also alter intertemporal saving incentives, which is particularly relevant for both AI and mitigation investment. To isolate the specific impact of AI, we therefore adopt a constant discount rate throughout our analysis. Specifically, we follow the comprehensive expert survey by \cite{drupp2018discounting} and use the mean rate recommended by economists as the benchmark parameter in the welfare function (Table \ref{tab_discount}). Recognizing that the discount rate remains one of the most debated parameters in climate policy evaluation \cite{pindyck2021we}, we perform sensitivity analyses using three alternative values: the mean rate suggested by philosophers \cite{nesje2023philosophers}, the constant pure rate of time preference from DICE-2016 \cite{nordhaus2018projections}, and the rate adopted by \cite{stern2007economics}. The latter two represent, respectively, relatively conservative and more stringent normative stances toward climate policy.

\begin{table}[htbp]
  \centering
  \caption{The discount rates used in DICE-AI}

  \begin{tabular}{lccp{20em}}
    \toprule
    & $\rho$ & $\eta$ & Source \\
    \midrule
    Benchmark & 0.011 & 1.35 & Mean value recommended by economists \cite{drupp2018discounting} \\
    \cmidrule{2-4}
    \multirow{3}{*}{Alternatives} & 0.009 & 1.4 & Mean value recommended by philosophers \cite{nesje2023philosophers} \\
    & 0.015 & 1.45 & DICE-2016 \cite{nordhaus2018projections} \\
    & 0.001 & 1 & Stern discounting \cite{stern2007economics} \\
    \bottomrule
  \end{tabular}
  \label{tab_discount}
\end{table}

\subsection{Economic frontier}

\subsubsection{AI sector production function}

In the DICE‑AI model, we introduce an intermediate AI production sector. This sector employs a standard Cobb–Douglas technology to combine AI‑specific capital $K_{AI,t}$ and labor $ L_{AI,t}$ to produce \emph{pseudo} AI output:
    \begin{equation}
    \label{eq:ai_y}
        y^{AI}_t=A_{AI,t} K_{AI,t}^\alpha L_{AI,t}^{1-\alpha}
    \end{equation}
where $A_{AI,t}$ represents the sector-specific technology level.

The AI‑sector capital stock evolves according to the perpetual‑inventory method with an assumed depreciation rate $\delta=10\%$:
\begin{equation}
    \label{eq:k_ai}
    K^{AI}_{t+1}=(1-\delta) K^{AI}_{t}+I^{AI}_{t}
\end{equation}
where the time-series data of AI investment $I^{AI}_{t}$ is drawn from Stanford University’s AI Index Report \cite{AIIndex2025}, which reports global corporate AI investment from 2013 to 2024. We adjust the reported figures \footnote{\label{fn_01}The data from AI index report capture annual corporate finance transactions involving privately held AI companies. They exclude publicly traded firms (e.g., large technology companies), companies’ internal expenditures such as R\&D or infrastructure investment, as well as government funding. For example, the European Union has launched the InvestAI initiative to mobilize €200 billion in artificial intelligence investment\cite{Bergeaud2025_EU_Productivity}. SecondTalent(2025) estimate that government funding accounts for around 40\% of total AI investment in China \cite{SecondTalent_ChinaAI_2025}, also the U.S. federal budget for AI R\&D is approximately USD 3 billion annually \cite{NITRD_AI_RnD_2025}. In addition, publicly listed firms also undertake  AI-related spending; for instance, Alphabet’s annual R\&D expenditure has been around USD 40 billion in recent years \cite{Alphabet10K2024}. Taking these omissions into account, we scale the date from Stanford University’s AI Index Report by a factor of 1.9 to construct a time series of total AI investment.} and express all values in constant 2019 PPP‑adjusted US dollars. 

Under a steady‑state approximation, the initial capital stock in 2013 is given by:
\begin{equation}
\label{eq:k_ai2013}
K^{AI}_{2013}\approx I^{AI}_{2013}/(g'+\delta)
\end{equation}
with $g'$  denoting the observed growth rate of AI investment between 2013 and 2014. Applying Eq.~(\ref{eq:k_ai}) recursively yields an AI‑sector capital stock of approximately USD 0.95 trillion in 2020.

For labor input in AI sector $L^{AI}_{t}$,  we assume for simplicity a fixed share of total labor force:
\begin{equation}
\label{eq:L_ai}
L^{AI}_{t}=v L^{Total}_{t}
\end{equation}
where total labor force $ L^{Total}$  is taken from the original DICE model and satisfies: 
\begin{equation}
\label{eq:Labor}
L^{AI}_{t}+L_{t}=L^{Total}_{t}
\end{equation}
with $L_{t}$ being labor in the final‑goods sector. The share parameter $v=0.3\%$  is calibrated using evidence from OECD statistics \cite{green2023supply} and the Stanford AI Index Report 2025 \cite{AIIndex2025}.

The capital‑income share of the AI sector $\alpha$ is calibrated by using the information and communication technology (ICT) sector as a proxy, following the ISIC Rev.4 classification \cite{UN2008ISIC}. Using national‑accounts data for the United States, the EU‑15, and China---treated as representative of the global economy---we compute the compensation‑of‑employees share in total gross value added. Data for the United States and the EU are obtained from the KLEM database \cite{EUKLEMS_INTANProd}, while data for China come from the National Bureau of Statistics \cite{NBS_IO_2020}.

\begin{figure}
    \centering
    \includegraphics[width=0.66\linewidth]{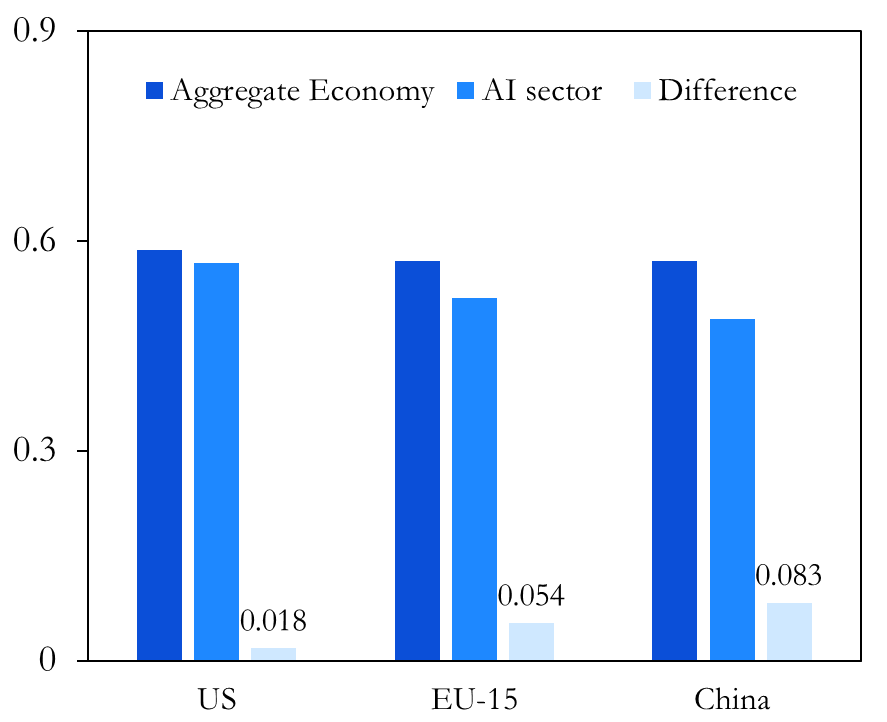}
    \caption{Labor income share for three representative regions in 2020 }
    \label{figD7}

\begin{tablenotes}
        \footnotesize
        \item \textbf{Note:} Author's calculation based on data from KLEM database \cite{EUKLEMS_INTANProd} and data from the National Bureau of Statistics of China \cite{NBS_IO_2020}.

    \end{tablenotes}

\end{figure}

As shown in Fig.~\ref{figD7}, the AI sector proxied by the ICT sector exhibits a lower labor‑income share than the aggregate economy, with the difference ranging from 0.018 to 0.083 across the three regions. This indicates that the AI sector is, on average, more capital‑intensive. Considering the economic weight of these regions and the aggregate labor‑income share of 0.7 in DICE‑2023, we set the AI‑sector labor‑income share to 0.66, which is 4 p.p. lower than the economy‑wide value.

Finally, the technology level of the AI sector, $A_{AI,t}$, is normalized to unity in the initial year and is assumed to grow at the same rate as labor productivity in the final‑goods sector.

\subsubsection{Final production sector: Mimicking the DICE-2023’s economy with DICE-AI}\label{ap_cal_final_sector}

The DICE-AI model incorporates an AI sector that competes with the final-goods sector for capital and labor. Consequently, less capital and labor remain available for final‑goods production. To preserve comparability with the total economic output of the original DICE‑2023 model, we adjust upwards the total factor productivity (TFP) of the final‑goods sector. We therefore recalibrate the final‑sector production function, Eq.~(\ref{eq:Y_t}), so that DICE‑AI with a fixed AI‑sector investment rate can replicate the economic frontier of DICE‑2023 in the absence of climate damages. Recall that final output is given by:
    \begin{equation}
    \label{eq:Y_t}
        Y_t=A_t(y_t^{AI}) \times K_t^{\gamma(y_t^{AI})}  L_t^{1-\gamma(y_t^{AI})}
    \end{equation}
where both the TFP level $A_t$ and the capital income share $\gamma$ depend on AI output. When calibrating the initial values of capital stock, labor and TFP, both $A_t(y_t^{AI})$ and $\gamma(y_t^{AI})$ are treated as constants, and the dependence on $y_t^{AI}$ can be ignored. The initial capital stock of the final sector is by definition equal to the difference between the total capital stock in DICE‑2023 and the AI‑sector capital stock calibrated earlier:
    \begin{equation}
    \label{eq:K_2020}
        K_{2020}= K^{total}_{2020}- K^{AI}_{2020}=295-0.95= 294.05 \text{ (trillions \$)} \notag
    \end{equation}
Similarly, labor in the final production sector is obtained by subtracting AI-sector employment from total labor in DICE-2023, with the initial level given by:
    \begin{equation}
    \label{eq:L_t}
        L_{2020}=L^{Total}_{2020}-L^{AI}_{2020}=7752.9-0.003×7752.9=7729.6  \text{ (millions)}  \notag
    \end{equation}

Finally, we recalibrate both the initial TFP level and its exogenous growth rate for the final sector. The initial TFP level is derived directly from Eq. (\ref{eq:Y_t}) using the derived initial capital stock and labor for the final sector---obtained from the calculations described above---together with the total output level taken from DICE‑2023. However, the TFP growth rate requires additional adjustment, because the economic projections in DICE‑2023 are based on expert surveys on long‑term growth that include scholars specializing in AI’s economic impact \cite{Christensen2018GrowthUncertainty} \cite{Rennert2021SCC}. To address this, we fix the AI‑sector investment rate at its observed 2020 level and fine‑tune the TFP path of the final sector so that total economic output in DICE‑AI matches that of DICE‑2023 exactly in 2100.

The initial value of AI investment is taken from Stanford University’s AI Index Report \cite{AIIndex2025} and adjusted according to Footnote \ref{fn_01}, yielding $ I^{AI}_{2020}=0.42$ trillion \$. With total output $ Y_{2020}=134.89$ trillion \$ in DICE-2023, this implies a fixed AI-sector investment rate of $\bar{s}_{AI}=0.42/134.89=0.31\%$.

\subsubsection{AI and economic growth}
Existing studies \cite{14015,trammell2023economic} suggest two principal channels through which AI reshapes economic growth, by enhancing total factor productivity (TFP) growth and fostering automation in tasks. AI‑driven TFP growth can be alternatively interpreted as the automation of research and development. Motivated by the observation that ideas—and hence growth—become harder to find over time \cite{bloom2020ideas}, we model AI‑induced TFP growth as:
\begin{equation}
\label{eq:g_tfp}
\frac{\dot{A_t}}{A_t}=g_{base,t}+\theta_{TFP}\times (y_t^{AI} -y_t^{AI,base}) \times A_t^{-\beta}
\end{equation}
where the term $ A_t^{-\beta}$ captures the diminishing marginal contribution of AI to TFP growth as the productivity level rises. Here, $ g_{base,t} $ is the exogenous TFP growth path of the final production sector (calibrated in Appendix \ref{ap_cal_final_sector}) . $ y_t^{AI,base}$ denotes the implicit AI output in DICE‑2023 (obtained by fixing the AI‑sector investment rate as in Appendix \ref{ap_cal_final_sector}), and $y_t^{AI} $
is the AI output level determined endogenously in the DICE‑AI model in equilibrium. The parameter $\theta_{TFP}$
 scales the spillover effect of AI output on TFP growth.

The factor‑income‑share impact, which reflects AI‑driven automation, is modeled as:
\begin{equation}
\label{eq:gamma}
\gamma(y_t^{AI})=\gamma_{asymp}-\gamma_{adj}\times e^{-\theta_{\gamma}\times (y_t^{AI} -y_t^{AI,base})}
\end{equation}
where $\gamma(y_t^{AI})$ is the capital income share in the final sector, $\gamma_{asymp}$ its the asymptotic upper bound, $\gamma_{adj}$  is chosen so that capital income share equals 0.3 when $y_t^{AI} =y_t^{AI,base}$. The parameter $\theta_{\gamma} $ governs  the speed at which the capital share converges toward $\gamma_{asymp}$. 

Quantitative estimates of AI‑driven growth in the existing literature are generally conservative, as they rely on historical data that largely treat AI as part of the broader ICT sector. The future impact of AI, however, remains highly uncertain. The rapid recent diffusion of generative AI suggests a plausible prospect in which AI triggers a new industrial revolution. To capture this uncertainty in calibrating the economic effects of AI---governed by Eqs. (\ref{eq:g_tfp}) and (\ref{eq:gamma})---we consider two benchmark prospects: a conservative ICT-like prospect and an aggressive industrial-revolution (IR)-like prospect. 

We begin with calibrating the parameters for the ICT-like prospect. Because both total factor productivity (TFP) growth and changes in factor income shares affect endogenous AI investment, we calibrate the two effects separately to avoid interference between them.

\begin{table}[htbp]
  \centering
  \caption{Survey on AI-driven total factor productivity growth}
  \setlength{\extrarowheight}{3pt}
  \begin{tabular}{p{9em}p{11.5em}p{3em}p{15em}}
    \toprule
    Sources & AI-driven TFP gains & Scope & Method  \\
    \midrule
    Filippucci \& Schief (2024) \cite{OECD_AI_2024} & Annual growth increases by 0.25-0.6 p.p over a 10-year horizon. & OECD & Estimate micro-level gains, map them to sectors, and aggregate economy-wide impacts using a multi-sector general equilibrium model with input-output linkages. \\[16pt]
    Aghion \& Bunel (2024) \cite{aghion2024ai} & Annual growth increases by 0.68 to 1.3 p.p over next decade & Global & Comparative analysis and theoretical modeling \\[16pt]
    Acemoglu (2025) \cite{acemoglu2025simple} & TFP gains over the next decade are 0.53\% or 0.66\%. & Global & Macroeconomic general equilibrium model \\[16pt]
    Rockall et al. (2025) \cite{Rockall2025AI} & Annual growth increases by 1.4\% or 1.7\% & Global & Structural macro model \\[16pt]
    Haskel (2025) \cite{Haskel2025_AI_Productivity} & Annual growth increases by 0.1 to 0.5 p.p & Global & A General-Purpose Technology Approach \\[16pt]
    Bergeaud (2024) \cite{Bergeaud2025_EU_Productivity} & A 1 p.p. increase in the IT capital share (within the total capital stock)  corresponds to a 6 \% rise in labor productivity.  & US \& EU & Ordinary Least Squares \\[16pt]
    Goldman Sachs (2023) \cite{GS_AI_Growth_2023} & Labor productivity growth increases by 1.5 p.p. over a decade. & US    & Decomposition-based accounting \\[16pt]
    Baily et al. (2023) \cite{Baily_AI_Productivity_2023} & Annual growth increases by 1.50 or 1.96 p.p & US    & Estimate approximately \\
    \bottomrule
  \end{tabular}%
  \label{tab_cal_eco}%

  \begin{tablenotes}
    \footnotesize
    \item \textbf{Note:} For analyses at the global scale, the existing quantitative evidence is largely derived from data on developed economies. Baily et al. (2023) \cite{Baily_AI_Productivity_2023} present future growth trajectories graphically, from which we infer the corresponding annual productivity gains attributable to AI.
  \end{tablenotes}
\end{table}%

For AI-driven TFP growth (Eq. \ref{eq:g_tfp}), we first hold the capital income share fixed at the DICE‑2023 value of 0.3. Following Bloom et al. (2020) \cite{bloom2020ideas}, who estimate the idea‑elasticity parameter $\beta$ across several U.S. sectors, we adopt their aggregate‑economy estimate $\beta=3.1$.  We then review existing studies on AI‑induced macroeconomic productivity gains (Table \ref{tab_cal_eco}), which report a wide range of estimates reflecting differing methodologies and time horizons. Adopting a conservative stance for the ICT‑like prospect, we calibrate  $\theta _{TFP}=4.98 $ such that, under endogenous AI investment in the economic‑frontier scenario of DICE‑AI, TFP increases by 0.265\% over the next decade (to 2030). This value corresponds to half of the central estimate by Acemoglu et al. (2025) \cite{acemoglu2025simple}, as it excludes AI‑driven changes in factor income shares and represents a deliberately cautious assessment.

To calibrate AI's impact on capital income share (Eq. \ref{eq:gamma}), in a similar vein, we first hold the TFP growth path of the final‑goods sector exogenously fixed at $g_{base,t}$, thereby excluding AI-driven TFP growth. Quantitative evidence on AI‑ and automation‑induced shifts in factor shares remains relatively limited.  Melina et al. (2024) \cite{IMF_2024_GenAI_Work} report that the capital income share in advanced economies rose by about 5.5 percentage points during the automation period from 1980 to 2014. Minniti et al. (2025) \cite{minniti2025ai} estimate that AI‑driven technological change may have reduced the labor share by 0.09–0.31 percentage points from a baseline of 52\%. Based on this evidence, we set a moderate asymptotic capital share $\gamma_{asymp}=0.34$ and choose $\gamma_{adj}=0.04$ so that the capital share equals 0.3 when $y_t^{AI}=y_t^{AI,base}$. The scale parameter $\theta_{\gamma}$ is calibrated to 0.283, with DICE-AI in its economic frontier yielding an increase of 0.01 percentage points in the capital income share by the end of the century under endogenous AI investment. We note that all these parameters are subject to considerable uncertainty.

For the industrial‑revolution (IR) prospect, quantitative evidence that directly estimates the magnitude of AI‑driven total factor productivity (TFP) growth or AI‑induced shifts in the capital income share remains scarce. We therefore employ a historical‑analogue approach, calibrating both effects by drawing on evidence from past industrial revolutions.

Bouscasse et al. (2025) \cite{bouscasse2025did} estimate productivity growth in England between 1250 and 1870 and show that the First Industrial Revolution (the Age of Steam, c. 1760–1840) raised productivity growth by a factor of 2.5, from about 2\% to 5\% per decade. Using the Maddison Database 2010 \cite{Maddison2010}, we conduct a similar analysis for the Second Industrial Revolution (the Technological Revolution, c. 1870–1914) and the Third Industrial Revolution (the Digital Revolution, post‑1970). Specifically, we compute the average annual growth rate of world per‑capita GDP before and after each revolution. The Second Industrial Revolution increased the annual growth rate by a factor of 2.24 (from 0.54\% to 1.20\%), and the Third Industrial Revolution raised it by a factor of 2.08 (from 1.20\% to 2.50\%) (Fig. \ref{ef04b}).

Guided by this historical evidence, we calibrate the IR prospect by setting $ \beta=2$ (an assumed value for heuristic calibration), $ \gamma_{asymp} =0.4$, $ \gamma_{adj}= 0.1$, and $\theta_{\gamma}=0.313$. Under this parameterization, capital deepening alone increases the capital income share by 5 p.p. by 2100, and the AI‑driven IR prospect raises the annual per capita GDP growth rate by approximately 2.3 times, consistent with the scale observed in the three previous industrial revolutions.

\begin{figure}
    \centering
    \includegraphics[width=0.66\linewidth]{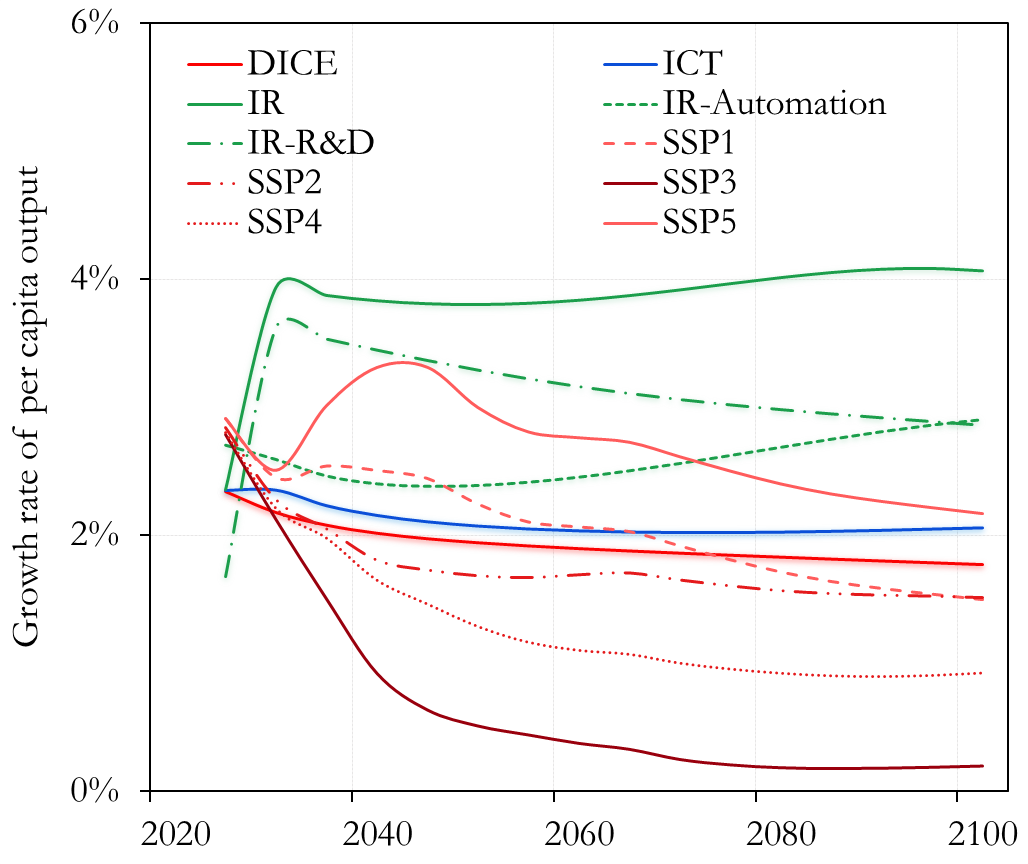}
    \caption{ DICE-AI's growth rate projections v.s. SSP
}
    \label{ef05}
    
    \vspace{2mm}
    \begin{minipage}{\textwidth}
        \footnotesize
        \textbf{Note:}  ICT projects a modest influence on long-run growth from AI, much like earlier information and communications technologies. IR posits a transformative economic impact sufficient to trigger a new industrial revolution. The Shared Socioeconomic Pathways (SSPs), adopted by the IPCC, provide standardized long-term projections of socioeconomic development in the absence of climate policy; data are sourced from the IIASA SSP Scenario Explorer  \cite{IIASA_SSPExplorer}.
    \end{minipage}
    
\end{figure}

We further present future economic growth trajectories in the economic frontier (excluding the climate impacts) for the ICT prospect, the IR prospect, and for IR scenarios that include only automation effects or only R\&D effects. hese outcomes are compared with DICE‑2023 and with historical growth during the three industrial revolutions (see Fig. \ref{ef04}).  Under the ICT prospect, AI yields modest but non‑negligible economic gains relative to DICE. In per‑capita GDP terms, AI raises output by about 4\% by mid‑century and 14\% by the end of the century (Fig.\ref{ef04a}). These magnitudes align with existing estimates in the literature. For example, Cerutti et al. (2025) \cite{cerutti2025global} report that AI‑induced productivity gains could increase global GDP by up to 4\% over the next decade, and Jones and Tonetti (2025) \cite{JonesTonetti_2025_Automation} find that AI and automation produce output gains of around 4\% by 2040. In contrast, under the IR prospect, the impact of AI on economic growth is comparable in scale to that of previous industrial revolutions (Fig \ref{ef04b}).

Finally, we compare the economic growth projections of DICE‑AI (both IR and ICT prospects) with those of the Shared Socioeconomic Pathways (SSPs) \cite{IIASA_SSPExplorer} used in IPCC AR6. As shown in Fig. \ref{ef05}, under the ICT prospect, the resulting per‑capita GDP growth rate lies between the SSP1, SSP2, and SSP5 trajectories, while under the IR prospect it exceeds the growth rates implied by all SSP pathways.

\subsection{Emission}\label{ap_cal_em}
\subsubsection{Capital-embodied carbon intensity in the AI sector}

\begin{table}[htbp]
  \centering
  \caption{Literature survey of data‑center emission estimates}
  \begin{tabular}{p{12em}p{6em}p{8em}p{6em}p{5em}}
    \toprule
    Sources & Scope & CO\textsubscript{2} emissions & Projection for 2030 & Embodied emissions included? \\
    \midrule
    Masanet et al. (2020) \cite{masanet2020recalibrating} 
      & Global (2018) & 208 TWh (95 MtCO\textsubscript{2}) & — & No \\[16pt]
    Shehabi et al. (2024) \cite{shehabi20242024} 
      & US (2018/2023) & 76/176 TWh (27/61 MtCO\textsubscript{2}) & 325–580 TWh (2028) & Partly \\[16pt]
    IEA (2025) \cite{IEA2025EnergyAI} 
      & Global (2020) & 269 TWh (123 MtCO\textsubscript{2}) & 669–1246 TWh & No \\[16pt]
    IEA (2023) \cite{IEA_2023_Tracking_Clean_Energy}   
      & Global (2020) & 330 MtCO\textsubscript{2} & — & Yes \\[16pt]
    McKinsey (2024) \cite{McKinsey_2025_Data_Centers} 
      & US (2023) & 147 TWh (51 MtCO\textsubscript{2}) & 606 TWh & No \\[16pt]
    Semianalysis (2024) \cite{SemiAnalysis_2024_AI_Datacenter} 
      & Global (2020) & 300 TWh (137 MtCO\textsubscript{2}) & 750–2150 TWh & No \\[16pt]
    Accenture (2025) \cite{Accenture_Sustainable_AI} 
      & Global (2022) & 38 Mt CO\textsubscript{2} & 491–756 TWh & No \\[16pt]
    World Bank (2021) \cite{WB_2024_Cloud_Infrastructure} 
      & Global (2022) & 500–700 TWh (229–321 MtCO\textsubscript{2}) & — & Partly \\[16pt]
    Xiao et al. (2025) \cite{xiao2025environmental} 
      & US (2024–2030) & 24–44 MtCO\textsubscript{2} & — & Partly \\
    \bottomrule
  \end{tabular}
  \label{tab_cal_AI_em}

  \begin{tablenotes}
    \footnotesize
    \item \textbf{Note:} Electricity consumption is converted to carbon emissions using the carbon intensity of electricity generation for the United States and the global average, as reported by the IEA (2021) \cite{IEA_2022_Emissions_Factors}. For SemiAnalysis (2024) \cite{SemiAnalysis_2024_AI_Datacenter}, numerical values are approximated based on figures in the report, as explicit values are not provided.
  \end{tablenotes}
\end{table}

We derive the initial carbon intensity of AI capital based on the range of estimates for data‑center emissions, which are commonly used in the literature as a proxy for AI‑sector emissions (Table \ref{tab_cal_AI_em}). For the year around 2020, reported global data‑center carbon emissions vary between 38 and 330 MtCO\textsubscript{2}, depending on system boundaries and the inclusion of embodied emissions. Given the global scope of our analysis and our inclusion of embodied emissions, we set initial‑year emissions of the AI sector to 220 MtCO\textsubscript{2}. The corresponding carbon intensity of AI capital is therefore given by:
\begin{equation}
\label{eq:sigma_kai0}
\sigma_{KAI, 2020 }=E_{AI, 2020}/K_{AI, 2020}=220/0.95=0.2316
\text{ (kg/\$)} \notag
\end{equation}

Next, we specify the rate of decline in the carbon intensity of AI capital. We assume that the AI sector decarbonizes faster than the aggregate economy before 2050, after which both converge to the same rate of decline. This assumption is supported by two lines of evidence.

First, emissions from data centers come largely from electricity consumption. The electricity sector has greater potential for rapid decarbonization than the broader economy, due to the expanding share of low‑carbon sources such as solar, wind, and hydropower. According to the IEA, the global carbon intensity of electricity generation has been declining sharply---with a record drop of about 3\% in 2024---and is projected to fall by an average of 3.6\% per year during 2025–2027 \cite{IEA_2025_Electricity}. In contrast, the annual decline in the carbon intensity of the overall economy has averaged only about 1.8\% in recent years (2015–2021). Looking ahead, the IEA's World Energy Outlook projects that under current policies, the global economy‑wide carbon intensity will decline at an average rate of only about 1\% per year between 2024 and 2050 \cite{IEA_2025_WEO_Dataset}.

Second, improvements in hardware and operational efficiency further reduce the energy intensity of data centers per unit of computation. Masanet et al. (2020) \cite{masanet2020recalibrating} estimate that advances in processor efficiency, reductions in idle power, and gains in storage density and efficiency have lowered the energy intensity of global data centers by an average of 20\% per year since 2010, substantially faster than efficiency improvements observed in most other energy‑intensive sectors. Consistently, assessments by Lawrence Berkeley National Laboratory highlight that higher server utilization and continued improvements in computing efficiency have been key to moderating data‑center energy growth over the past decade \cite{shehabi20242024}.

Building on this combined evidence, we assume a faster decline in carbon intensity for the AI sector than the aggregate economy before 2050, with both converging to the same rate thereafter. Specifically, the decline rate of the AI sector in 2020 is set to be 3.4 times that of the final production sector in DICE-2023. This factor is assumed to decrease gradually over time, falling to 3.0 by 2025, 2.6 by 2030, 2.2 by 2035, 1.8 by 2040, and 1.4 by 2045. From 2050 onward, the decline rates of the AI sector and the aggregate economy are identical.

Moreover, to capture uncertainty in the decline rate of AI-sector carbon intensity, we conduct a sensitivity analysis by first specifying a distribution for this parameter. Specifically, we determine its lower and upper bounds and then assume a normal distribution, with the baseline value set at the distribution mean. For the lower bound, the evidence above suggests that the decline rate of AI-sector carbon intensity is about 1.6 times that of the overall economy (3\%/1.8\%). For the upper bound, we draw on \cite{masanet2020recalibrating}, who report an average annual reduction of about 20\% in energy use per unit of computation since 2010. According to \cite{IEA2025Efficiency}, the global annul rate of energy-efficiency improvement is around 1.75\% since 2010. Because the estimate in \cite{masanet2020recalibrating} captures only computing-related efficiency gains, we adjust it downward to 7.83\% \footnote{\label{fn_05}  According to Table \ref{tab_cal_AI_em} and estimates from IEA (2023, 2025), embodied energy accounts for about two-thirds of total energy consumption (\cite{IEA2025EnergyAI}; \cite{IEA_2023_Tracking_Clean_Energy}). We assume that this component improves at the global average energy-efficiency rate of 1.75\%. The combined energy-efficiency improvement rate for data centres is therefore 1.75\%×23+20\%×13=7.83\%} to account for embodied energy and its efficiency. This implies an upper-bound ratio of 4.5 (7.83\%/1.75\%) for the decline rate of AI-sector carbon intensity relative to that of the overall economy. 

To center the normal distribution at our baseline calibration of 3.4, we calibrate the standard deviation using the more distant bound. Given the remaining uncertainty, we assume that the lower–upper bound range covers 90\% of the distribution. Thus,

\begin{equation} 
\label{eq:sd_decline}
SD_{decline }=(3.4-1.6)/1.645=1.094 \end{equation}

The resulting values at the 2.5th, 25th, 50th, 75th, and 97.5th percentiles are 1.26, 2.66, 3.40, 4.14, and 5.54, respectively. These five values are used in the sensitivity analysis, while the calibrated normal distribution is used for the Monte Carlo simulations.

\subsubsection{Capital-embodied carbon intensity in the final sector}

\begin{figure}[tb]
    \centering
    \begin{subfigure}[b]{0.49\textwidth}
        \centering
        \includegraphics[width=\textwidth]{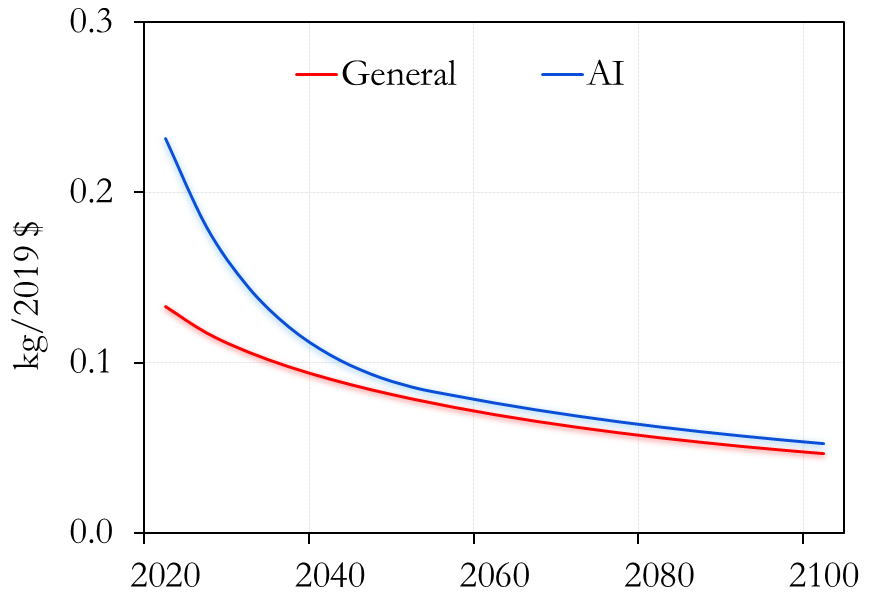}
    \caption{Capital-embodied carbon intensity}
        \label{ef6a}
    \end{subfigure}
    \hfill
    \begin{subfigure}[b]{0.49\textwidth}
        \centering
        \includegraphics[width=\textwidth]{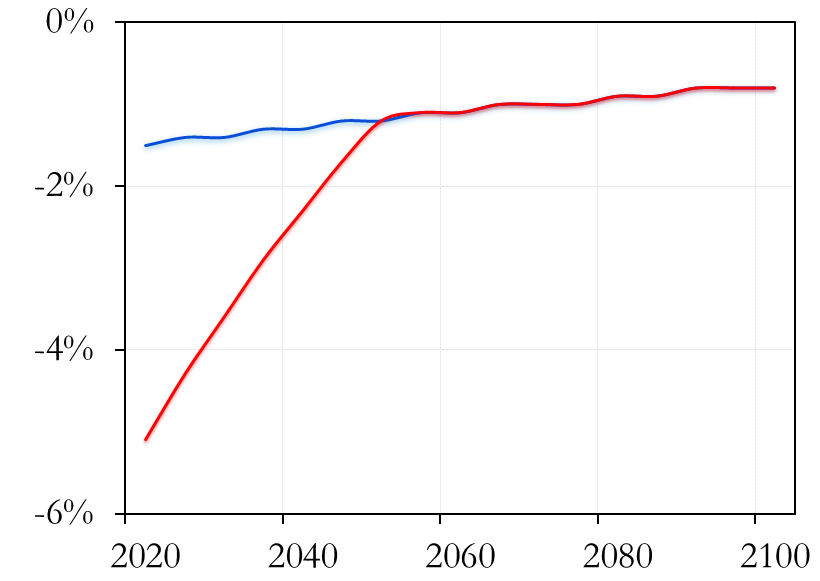}
        \caption{Annual growth of carbon intensity}
        \label{ef6b}
    \end{subfigure}

\vspace{0.5 em}
  
    \caption{Capital-embodied carbon intensity in DICE-AI}
    \label{ef06}
    
\end{figure}

Given the time series of AI‑sector emissions and the corresponding trajectories of both types of capital in the economic frontier (without climate impacts), the carbon intensity of general capital is computed as::
\begin{equation} 
\label{eq:sigma_k}
\sigma_{K, t }=(E_{total,t}-E_{AI,t})/(K_{total,t}-K_{AI,t}) \end{equation}
where $ E_{total,t}$  and $ K_{total,t}$ are taken from the original DICE-2023 model. Fig. \ref{ef06} presents the calibrated capital-embodied carbon intensity and its annual growth rate for both general capital and AI capital in our DICE-AI model. We perform this calibration under the economic‑frontier scenario to avoid a circular identification problem. In scenarios that include the full climate module (with damages and abatement costs), AI‑sector investment----and hence AI capital---is endogenously determined and varies with the very parameters that govern carbon intensity. Because those parameters are themselves calibrated as functions of $K_{AI}$ such a setting would create a circular dependence and prevent a well‑defined calibration. By contrast, in the economic frontier, the path of $K_{AI}$ remains stable regardless of the emissions‑related parameters, allowing a consistent and unambiguous calibration.

\subsubsection{AI-powered emission conservation}

To model the emissions‑reduction effect of AI, we introduce a multiplicative scaling factor applied to the aggregate of capital-embodied emissions. Specifically, $\Omega(y^{AI}_t )$ captures the share of emissions reduced as AI output rises, such that the factor $1-\Omega(y^{AI}_t )$ represents the fraction of emissions remaining after AI‑driven abatement. The factor is specified as:
\begin{equation}
\label{eq:omega}
(1-\Omega(y_t^{AI}))= \psi_{asymp}+\psi_{adj}\times e^ {-\theta_{sav} \times (y_t^{AI} -y_t^{AI,base})}
\end{equation}
where $\psi_{asymp}+\psi_{adj}=1$. By construction, when $y_t^{AI} =y_t^{AI,base}$, the factor equals unity, so no abatement occurs. As AI output grows, the factor converges to $\psi_{asymp}$, the asymptotic lower bound that represents AI’s maximum mitigation potential. The parameter $ \theta_{sav}$ governs the speed at which this mitigation effect materialises with increasing AI adoption.

Because direct estimates of $\psi_{adj}$ are not available in the literature, we calibrate it based on a review of related studies (Table \ref{tab_cal_AI_abate}). We set $\psi_{adj}=0.1$, which by construction implies $\psi_{asymp}=0.9$. The adjustment parameter $ \theta_{sav}  $  is then calibrated to 4.315, so that under a business‑as‑usual (BAU) scenario AI adoption reduces global emissions by approximately 3 GtCO\textsubscript{2} in 2035. This 3 GtCO\textsubscript{2} reduction lies midway between the estimates reported by Stern et al. (2025) \cite{stern2025green} and the IEA (2025) \cite{IEA2025EnergyAI}. The BAU scenario is used because both references already incorporate existing abatement policies in their quantitative estimates.

\begin{table}[htbp]
  \centering
  \caption{Literature survey of AI's abatement potentials}
    \begin{tabular}{p{6em}p{12em}p{12em}p{12em}}
    \toprule
    Sources & Describe  & Channel & Method  \\
    \midrule
    Knittel \& Stolper (2025) \cite{knittel2025using} & AI reduces household electricity use by an average of 9 kWh per month. & Causal forests with selective targeting improve the effectiveness of randomized behavior  nudges for household energy conservation. & Difference-in-Differences regression \\[16pt]
    Stern et al. (2025) \cite{stern2025green} & AI reduces annual global BAU emissions in covered sectors by 3.2–5.4 GtCO\textsubscript{2}e by 2035 (13.7 to 24.1\%) & To transform complex systems, accelerate technological discovery \& resource efficiency, enable behavioural change, model climate systems \& policy interventions, and strengthen adaptation \& resilience & Bottom-Up Quantitative Estimation \\[16pt]
    Hua et al. (2025) \cite{hua2025artificial} & AI can reduce emissions by up to 15\%  & Forecasting, management and real-time monitoring of carbon emissions & Literature review \\[16pt]
    IEA (2025) \cite{IEA2025EnergyAI} & AI could reduce CO\textsubscript{2} emissions by about 1,400 Mt in 2035, excluding potential AI-driven breakthroughs. & To enable diverse optimizations that reduce emissions & A hybrid top-down–bottom-up, data-driven framework \\[16pt]
    Ding et al. (2024) \cite{ding2024potential} & In U.S. office buildings, AI adoption could reduce emissions by 8–19\% by 2050; combined with energy policies and low-carbon power, reductions could reach ~90\% & To expand high-efficiency and net-zero building penetration & Bottom-up building energy modeling approach combined with scenario analysis \\[16pt]
    Wang et al. (2024) \cite{wang2024ecological} & A 1\% increase in AI reduces carbon emissions by 0.0081\%. & To improve production efficiency, optimize processes, and support environmental protection and energy transitions & OLS: System Generalized Method of Moments (SYS-GMM) and Dynamic Panel Threshold Models (DPTM) \\[16pt]
    Li et al. (2023) \cite{li2023carbon} & In China, AI reduced carbon emissions from 5G mobile networks by 20.90 ± 0.98 Mt in 2023. & To leverage collaborative deep reinforcement learning and graph neural networks to coordinate 5G cell operations & Data-driven, simulation-based operational carbon accounting approach \\[16pt]
    Wolf et al. (2023) \cite{wolf2022potential} & Autonomous AI encounters can reduce GHG emissions by up to 80\% relative to in-person specialist visits & To reduce travel-associated GHGs through enhanced telemedicine & Real case: bottom-up, process-based carbon footprint analysis with counterfactual scenarios. \\
    \bottomrule
    \end{tabular}%
  \label{tab_cal_AI_abate}%
\end{table}%
Similarly, for the sensitivity analysis of this parameter, we first specify a normal distribution by determining its lower and upper bounds, with the baseline calibration of $\psi_{adj}=0.1$ set as the mean. Based on the literature reviewed in Table \ref{tab_cal_AI_abate}, the upper bound is 24\% (\cite{stern2025green}) and the lower bound is 4\% (\cite{IEA2025EnergyAI}). To center the distribution at the baseline value, we use the more distant bound to calibrate the standard deviation. To rule out negative values, we further reset the lower bound to zero and assume that the resulting range covers the 95\% confidence interval. The standard deviation is therefore given by:

\begin{equation} 
\label{eq:sd_abatement}
SD_{abate }=(0.1-0)/1.96=0.05 \end{equation}
The resulting values of  $\psi_{adj} $
at the 2.5th, 25th, 50th, 75th, and 97.5th percentiles are 0.00, 0.07, 0.10, 0.13, and 0.20, respectively. These five values are used in the sensitivity analysis, while the calibrated normal distribution with mean 0.10 and standard deviation 0.05 is used for the Monte Carlo simulations. 

\subsection{Damage}
\subsubsection{Adaptation function}

AI adaptation is incorporated directly into the climate damage function. The original DICE-2023 damage function is specified as
$D(T_t)=a_1T_t^{a_2}$, with $a_1=0.003467$, and $a_2=2$ such that damages are output losses of 3.1\% at 3\degree C warming and 7.0\% loss at 4.5\degree C warming \cite{barrage2024policies}. We extend this specification to include the adaptation effect of accumulated AI capital:
\begin{equation}
\label{eq:damage}
D(T_t,K_{AI,t})=\left[(1-a_3)+a_3\left(\frac{K_{AI,0}}{K_{AI,t}}\right)^{a_4}\right]\times a_1T_t^{a_2}
\end{equation}
This specification implies that, as AI capital accumulates, it asymptotically alleviates a fraction  $ a_3 $  of climate damage, leaving a residual fraction $1-a_3$. To calibrate $a_3$, we proceed in three steps. We first adopt the sectoral climate damage estimates reported by Tol (2024) \cite{Tol2024ClimateMeta}, ensuring that their aggregate effect is consistent with the DICE-2023 benchmark of a 3.12\% loss at 3\degree C of warming. Second, we assume that an infinitely large stock of AI capital can offset half of the climate damages in selected, adaptation-sensitive sectors, while leaving other sectors unaffected \cite{forouheshfar2025enhancing}. As summarized in Table \ref{tab_cal_AI_adapt}, this assumption results in an aggregate remaining climate damage of 1.81\% at 3\degree C of warming after accounting for AI adaptation, implying that $1-a_3=1.81/3.12$ such that $a_3=0.42$.

\begin{table}[htbp]
  \centering
  \caption{Sectoral climate damages and AI-powered adaptation potentials}
    \begin{tabular}{lcccc}
    \toprule
          & Share & \% GDP & Max impact of AI & \% GDP \\
    \midrule
    Agriculture & 0.085855 & -0.26787 & 0.5   & -0.13393 \\
    Forestry & 0.004259 & -0.01329 & 1     & -0.01329 \\
    Energy & 0.043571 & -0.13594 & 0.5   & -0.06797 \\
    Water & 0.038483 & -0.12007 & 0.5   & -0.06003 \\
    Tourism & 0.033904 & -0.10578 & 1     & -0.10578 \\
    Other markets & 0.242147 & -0.7555 & 0.5   & -0.37775 \\
    Coastal defence & 0.010117 & -0.03157 & 0.5   & -0.01578 \\
    Dryland & 0.023805 & -0.07427 & 1     & -0.07427 \\
    Wetland & 0.062138 & -0.19387 & 1     & -0.19387 \\
    Ecosystem & 0.073789 & -0.23022 & 0.5   & -0.11511 \\
    Health & 0.176982 & -0.55218 & 0.5   & -0.27609 \\
    Air pollution & 0.028585 & -0.08919 & 0.5   & -0.04459 \\
    Time use & -0.10632 & 0.331724 & 1     & 0.331724 \\
    Settlements & 0.062327 & -0.19446 & 0.5   & -0.09723 \\
    Catastrophe & 0.077528 & -0.24189 & 0.5   & -0.12094 \\
    Migration & 0.02017 & -0.06293 & 1     & -0.06293 \\
    Amenity & 0.122662 & -0.38271 & 1     & -0.38271 \\
    \midrule
    Total &       & -3.12 &       & -1.81056 \\
    \bottomrule
    \end{tabular}%
  \label{tab_cal_AI_adapt}%
\end{table}%

Last, we assume that by 2050, AI achieves half of its maximum potential adaptation effect, measured relative to the gap between its initial capital level (2020) and its asymptotic (infinite-capital) benchmark. Under the BAU scenario, AI capital increases from 0.95 trillion in 2020 to 51.50 trillion in 2050. Accordingly, $(0.95/51.5)^{a_4}=0.5$. Thus, $a_4=\ln(0.5)/\ln(0.95/51.5)=0.1736$.

Note that when are calibrating the DICE-AI model to mimic the DICE-2023 model, we explicitly incorporate an AI sector with a fixed path of AI-sector investment rates but retain the original DICE damage function without adaptation. This reflects the assumption that climate damages in DICE-2023 are not adjusted to account for AI-driven adaptation effects.  In the full DICE‑AI model, however, climate damages are modified by Eq. \ref{eq:damage}, which explicitly accounts for adaptation enabled by AI capital.

Similarly, to conduct the sensitivity analysis for AI-powered adaptation, we first determine the lower and upper bounds of , with the baseline value set at the distribution mean. Our calibration is based on the systematic review by \cite{forouheshfar2025enhancing}, which identifies sectors in which AI can enhance resilience to climate change. For the lower-bound case, we exclude sectors that are less closely related to those identified by \cite{forouheshfar2025enhancing}
from the set of adaptation-sensitive sectors assumed in the baseline calibration. This yields an aggregate residual climate damage of 2.33\% at  warming after accounting for AI adaptation, implying that,  $1-a_3=2.33/3.12$  such that $a_3=0.253$.

For the upper-bound case of AI-driven adaptation, we additionally include sectors that are somewhat related to those identified by \cite{forouheshfar2025enhancing}
but were not treated as adaptation-sensitive in the baseline calibration. This yields aggregate residual climate damage of 1.56\% at  warming after accounting for AI adaptation, implying and thus .
  $1-a_3=1.56/3.12$  such that $a_3=0.5$.

To center the normal distribution at our baseline calibration of  $a_3=0.42$ , we calibrate the standard deviation using the more distant bound. Given that this range is grounded in the literature, we assume that it covers 95\% of the distribution. Therefore, 

\begin{equation} 
\label{eq:sd_adapt}
SD_{decline }=(0.42-0.253)/1.96=0.085 \end{equation}

The resulting values of $a_3$ at the 2.5th, 25th, 50th, 75th, and 97.5th percentiles are 0.253, 0.362, 0.420, 0.477, and 0.586, respectively. These five values are used in the sensitivity analysis, while the calibrated normal distribution is used for the Monte Carlo simulations. 

\subsubsection{Alternative damage estimates for sensitivity analysis}

In DICE-2023, the damage function is constructed from three components: a literature-synthesis estimate of global aggregate warming impacts, implying a 1.62\% GDP-equivalent loss at 3\degree C warming; an adjustment for tipping points, adding a further 1\% output loss at 3 °C; and a judgment-based allowance of 0.5\% to account for omitted impacts and residual uncertainty. Given that the damage function remains one of the most thorniest issues in climate-change economics, we consider alternative damage estimates in sensitivity analyses.

More specifically, we draw on \cite{Tol2024ClimateMeta}
, which also serves as the basis for our calibration of AI-driven adaptation. The author provides an updated meta-analysis of the economic impacts of climate change, and estimates seven alternative damage functions using 69 estimates drawn from studies on the total economic impact of climate change. The function parameters are calibrated by minimizing the weighted sum of squared deviations between the model predictions and the primary estimates, assigning equal weight to all estimates within a given study and a total weight of one to each study. We adopt, for sensitivity analysis, the function with the highest likelihood among the seven alternatives as the first damage specification,
\begin{equation}
\label{eq:damage1}
    D(T_t)=0.01\times ( 0.45\times T_{t}+0.082\times T_{t}^{2})
\end{equation}

 This specification implies GDP losses of around 2.1\% under a 3°C increase in global mean surface temperature relative to the preindustrial period, which is lower than the corresponding estimate in DICE, where damages reach 3.12\% at 3°C warming. This discrepancy stems primarily from differences in literature search, study selection, weighting schemes, and the treatment of methodological heterogeneity across studies.

In addition, \cite{Tol2024ClimateMeta} estimates a fitted global impact function based on studies linking economic growth to temperature levels, which we use as the second alternative damage function in the sensitivity analysis,

\begin{equation}
\label{eq:damage2}
    D(T_t)= 0.002006\times T_{t}+0.000023\times T_{t}^{2}
\end{equation}
implying an approximately 0.6\% reduction in the annual growth rate of global economic output under 3°C of global warming. The growth effect generates substantially larger cumulative damages than the level effect.

Based on previous calibration Eq. \ref{eq:damage}, when accounting for AI-driven adaptation effects, Eq. \ref{eq:damage1} and \ref{eq:damage2} are modified as follows:

\begin{align}
\label{eq:damage1ca}
D(T_t,K_{AI,t})&=\left[0.58+0.42\times\left(\frac{0.95}{K_{AI,t}}\right)^{0.1736}\right]\times0.01\times ( 0.45\times T_{t}+0.082\times T_{t}^{2})\\
\label{eq:damage2ca}
D(T_t,K_{AI,t})&=\left[0.58+0.42\times\left(\frac{0.95}{K_{AI,t}}\right)^{0.1736}\right]\times (0.002006\times T_{t}+0.000023\times T_{t}^{2})
\end{align}

In the sensitivity analysis, we apply Eq. \ref{eq:damage1} and  Eq. \ref{eq:damage2}
to DICE-2023, while  Eq. \ref{eq:damage1ca} and Eq. \ref{eq:damage2ca} are applied to DICE-AI, with the growth-rate effect incorporated into the TFP growth equation.

\subsection{Carbon cycle and climate modules}

The climate module and its parameters in DICE-AI follow DICE-2023, including the carbon cycle, radiative forcing, and climate dynamics. CO\textsubscript{2} emissions enter through the carbon cycle before affecting radiative forcing, while non-CO\textsubscript{2} greenhouse gases are linked directly to radiative forcing and short-circuit the atmospheric chemistry.
    \begin{equation}
    \label{eq:R}
   \Delta R^i (t+1)=\xi_i E(t)-\left[ \frac{R^i(t)}{{\alpha(t) }\tau_i}\right] , i=1,...,4\end{equation}
   \begin{equation}
   \label{eq:mat}
   MAT(t)-MAT(1765)=\sum_{i=1}^4 R^i(t)\end{equation}
   \begin{equation}
   \label{eq:cacc}
 Cacc(t)=\sum_{v=1765}^t E(v)-[MAT(t)-MAT(1765)] \end{equation}

  Eq. \ref{eq:R} describes the four-reservoir carbon cycle, where  $R^i (t) $ denotes the carbon stock in reservoir  $i $ and $ E(t)    $  represents CO\textsubscript{2} emissions. The structural parameter  $ \alpha(t)   $ captures the fraction of cumulative CO\textsubscript{2} emissions remaining in the atmosphere as total emissions increase. Following DICE-2023, the fractions of emissions entering the four reservoirs are $\xi_i=(0.2173, 0.2240, 0.2824,0.2824 )$
, with corresponding decay time constants 
$\tau_i=(1,000,000, 394.4, 36.53, 4.304)$ years. Eq. \ref{eq:mat} aggregates the four reservoirs to obtain atmospheric CO\textsubscript{2}, $ MAT(t)$. Eq. \ref{eq:cacc} defines accumulated CO\textsubscript{2} in non-atmospheric sinks, $Cacc(t)$. 

\begin{equation}
\label{eq:irf}
iIRF100(t)=\varsigma_0 + \varsigma_C Cacc(t) + \varsigma_T T_{AT}(t)\end{equation}

\begin{equation}
\label{eq:irf1}
iIRF100(t)=\sum_{i=1}^4 {\alpha(t)} \xi_i \tau_i \left[1-e^{\frac{-100}{{\alpha(t)}\tau_i}}\right] \end{equation}

Eq. \ref{eq:irf} computes the 100-year integrated impulse response function, $i IRF100(t)$, where $T_{AT}(t)$denotes global mean surface temperature. The associated parameters are set as follows: the pre-industrial impulse response time is $\varsigma_0=32.4$ years; the IRF increases by 0.019 years per GtC of cumulative carbon uptake ($ \varsigma_C =0.019 $); and by 4.165 years per kelvin of warming  ($\varsigma_T =4.165 $). Eq. \ref{eq:irf1} implicitly defines the saturation parameter by equating Eq \ref{eq:irf} and \ref{eq:irf1}. Note that values of $R, Cacc,E $ are all zero in 1765.
        \begin{equation}
        \label{eq:force}
            F(t)=F_{CO_2 2x}\times\log_2 \left[\frac{MAT(t)}{MAT(1765)}\right]+F_{ABATE}(t)+F_{FX}(t)
        \end{equation}
        \begin{equation}
        \label{eq:tbox1}
          Tbox_1(t+1)=Tbox_1(t)e^{\frac{-5}{d_1}}+teq_1 F(t+1) \left(1-e^{\frac{-5}{d_1}}\right)
        \end{equation}
                   \begin{equation}
                   \label{eq:tbox2}
         Tbox_2(t+1)=Tbox_2(t)e^{\frac{-5}{d_2}}+teq_2F(t+1)\left(1-e^{\frac{-5}{d_2}}\right) 
        \end{equation}
        \begin{equation}
        \label{eq:tat}
       T_{AT}(t)=Tbox_1(t)+Tbox_2(t)
        \end{equation}
        \begin{equation}
        \label{eq:ecs}
      ECS=F_{CO_2 2x}\times(teq_1+teq_2)=3^\circ C
        \end{equation}

The remaining climate equations describe radiative forcing and global mean temperature. Greenhouse gas accumulation increases surface temperatures through higher radiative forcing, with the forcing–concentration relationship derived from empirical evidence and climate models. In Eq. \ref{eq:force}, $F(t)$ denotes the change in total anthropogenic radiative forcing since 1765 from CO\textsubscript{2} and other GHGs; $F_{EX}(t)$ represents exogenous forcing from non-abatable sources; and $F_{ABATE}(t)$ captures forcing from abatable non-CO\textsubscript{2} GHGs. Pre-industrial carbon stocks in 1765 are taken as the equilibrium baseline. 

Moreover, the climate module in Eqs. \ref{eq:tbox1}–\ref{eq:tat} adopts a two-box temperature model to represent the response to radiative forcing. Global mean surface temperature, $T_{AT}(t)$ is calculated as the sum of two components, $Tbox1$ and $Tbox2$. Parameters $d_1$ and $d_2$ denote the time lags (in years) of the two temperature boxes and are set to 236 and 4.07, respectively. Parameters $teq_1$ and $ teq_2$ represent the corresponding diffusion rates (in m\textsuperscript{2}·K/W), set to 0.324 and 0.44. Additionally, the equilibrium climate sensitivity is defined in Eq. \ref{eq:ecs}. 

\printbibliography[title=Supplementary References]
\end{refsection}



\end{appendices}



\end{document}